\documentclass[iop,numberedappendix,twocolappendix]{emulateapj}
\usepackage{amsmath}
\usepackage{amssymb}
\usepackage{epstopdf}
\usepackage{graphicx}
\usepackage{float}
\usepackage{natbib}
\usepackage{footnote}
\usepackage{algorithm2e,algorithmic}
\usepackage{bbold}
\usepackage{paralist}

\slugcomment{Submitted to ApJ}

\shorttitle{Proper Image Subtraction}
\shortauthors{Zackay, Ofek and Gal-Yam}

\begin{document}


\title{Proper image subtraction - optimal transient detection, photometry and hypothesis testing}


\author{Barak Zackay\altaffilmark{}, Eran O. Ofek\altaffilmark{} and Avishay Gal-Yam\altaffilmark{}}
\affil{Benoziyo Center for Astrophysics, Weizmann Institute of Science, 76100 Rehovot, Israel}

\email{bzackay@gmail.com}
\email{eran.ofek@weizmann.ac.il}




\begin{abstract}
Transient detection and flux measurement via image subtraction stand at the base of time domain astronomy.
Due to the varying seeing conditions, the image subtraction process
is non-trivial, and existing solutions suffer from a variety of problems. 
Starting from basic statistical principles, we develop the optimal statistic for transient detection,
flux measurement and any image-difference hypothesis testing.
We derive a closed-form statistic that:
(i) Is mathematically proven to be
the optimal transient detection statistic in the limit of background-dominated noise;
(ii) Is numerically stable;
(iii) For accurately registered, adequately sampled images,
does~not leave subtraction or deconvolution artifacts;
(iv) Allows automatic transient detection to the theoretical sensitivity limit by providing credible detection
significance;
(v) Has uncorrelated white noise;
(vi) Is a sufficient statistic for any further statistical test on the difference image, and
in particular, allows to distinguish particle hits and other image artifacts from real transients;
(vii) Is symmetric to the exchange of the new and reference images;
(viii) Is at least an order of magnitude faster to compute than some popular methods;
and (ix) Is straightforward to implement.
Furthermore, we present extensions of this method that
make it resilient to registration errors, color-refraction errors, and any noise source that can be modelled.
In addition, we show that the optimal way to prepare a reference image is the proper image coaddition presented in Zackay \& Ofek (2015b).
We demonstrate this method on simulated data and real observations from the Palomar Transient Factory data release 2.
We provide an implementation of this algorithm in MATLAB and Python.

\end{abstract}



\section{Introduction}\label{sec:Introduction}
Detection of previously-unknown transient sources is at the base of many fields of astronomy.
Examples include: the searches for supernovae, microlensing events and light echos.
To remove a constant complex background, it is useful to perform digital image subtraction,
a problem that has proven to be hard to tackle, with several suggested solutions
(e.g., \citealp{Phillips95,AL98,Bramich,Gal-Yam08,Yuan08}).
Probably the most popular algorithms are by \cite{AL98}
and \cite{Bramich}.

Current methods have several problems and limitations.
An important difficulty in image subtraction is that the point spread function (PSF)
of images taken from the ground
is varying\footnote{Sometimes this is relevant also for space-based observation.}.
In some cases, the subtraction is based on a numerically unstable process
(deconvolution) that may generate subtraction artifacts.
Combined with ill-defined error propagation\footnote{There are several reasons why the current methods propagate
the errors incorrectly. One reason is that convolution generates correlated noise, which is typically ignored. Second is that usually the errors in the reference image are not projected correctly.},
it is difficult to decide if a transient candidate is real or rather due to a subtraction artifact.
Finally, there is no proof that any of the methods we are currently using is optimal.
As we will show in this paper, none of these algorithms is optimal.
One hint for this is that some of these methods are not symmetric to exchange of the reference image and the new image, while the problem is symmetric.
Another hint is that none of the methods defines the matched filter that one should use in order to detect transients in the difference image.

In the \cite{AL98} and \cite{Bramich} class of solutions, a complex inversion problem needs to be solved.
This inversion problem can be regarded as a regularization effort (e.g., \citealp{Becker2012}) on the partial deconvolution
done by \citet{Phillips95}.
Apart from being computationally slow, this inversion problem
is in itself an effective deconvolution,
and the numerical instability of the deconvolution process cannot be swept under the rug.
These algorithms explore the trade off between ringing artifacts in the
subtraction image, that are due to the effective division in the Fourier plane,
and residuals from the constant-in-time sky
that are due to a failure of equalizing the PSFs of the reference and the new images.
For example, if the PSF of the new image is sharper than the PSF of the reference in some axis, then these methods find no good solutions leading to multiple image artifacts.

These artifacts,
along with residuals caused by registration errors,
appear as false positive signals that hinder the automatic detection of transients.
The current state of the art solution to this problem is to train a machine-learning algorithm
(e.g., \citealp{ML1,ML2,ML3}) to filter most of the artifacts and reduce the number of false positives to the minimum.
However, this solution is partial and human scanners
are required to sift through all remaining candidate detections and decide which is real and which is not (e.g., \citealp{Gal-Yam2011,Smith2011}).

This elaborate process can undermine the successful operation of transient searches in many ways.
First, employing many human scanners can be cumbersome and expensive.
Current surveys are spending
considerable manpower on candidate sifting (e.g., PTF).
Without further dramatic improvement,
this use of human scanners is unscalable, and is unfeasible for future surveys like ZTF \citep{Bellm2015} and LSST \citep{LSST}.
Second, having humans in the loop introduces a time delay
in the transient detection.
This can compromise science cases in which it is of utmost importance to make rapid follow-up observations
of new transients (e.g., \citealp{Cenko2013}, \citealp{Gal-Yam2014}, \citealp{Cenko2015}). 
Moreover, our experience is that at least some
machine learning algorithms throw away real obvious
transients.
Furthermore, the human scanning step makes it difficult
to estimate the completeness of transient surveys
as human scanners are difficult to properly simulate.
Another problem is that even human scanners can be unsure if a transient is
real or an artifact,
and many surveys adopt the methodology of accepting only candidates that are persistent
in two or more consecutive observations\footnote{This step is also required for unknown minor planet
identification.} (e.g., \citealp{Gal-Yam2011}, \citealp{LSQ}).
This methodology trades the survey speed with the increased credibility of the candidates,
and causes an additional time delay in transient detection.
%
Last, human scanning makes it difficult to detect transients at the faintest limit,
as it is hard for humans to objectively quantify the false alarm probability.

In this paper, we present a closed-form solution for image subtraction in general,
and transient detection in particular.
Starting with the most basic statistical principles, we solve the problem of transient detection under the assumption that both the reference and the new images have white Gaussian noise (e.g., the background-noise- or read-noise-dominated limit).
We then characterize the statistical behavior of our 
closed-form transient detection statistic under the influence of source noise and astrometric errors. Based on this analysis, we then construct a correction term to the transient detection statistic that
prevents false positive detections in the vicinity of bright objects.
Our solution is always numerically stable, is trivial to implement and
analyze, and is significantly faster computationally than the popular algorithms
(e.g., \citealp{AL98,Bramich}).
We extend the transient detection statistic to the situation of multiple references, and show that the optimal reference image for image subtraction is the proper coaddition image given in Zackay \& Ofek (2015b).
Finally, we show that the transient detection statistic is the maximal S/N
estimator for transient flux measurement in the background-dominated noise limit.

We further develop the optimal transient detection statistic into a difference image statistic that has white noise. Then, we show that any statistical measurement or decision on the data can be performed optimally and intuitively on this difference image, which we call the
{\it proper image subtraction statistic}.
This image has many good qualities such as: in the case of no difference between the reference and the new image, it has expectancy zero everywhere and
uncorrelated additive Gaussian noise.
It has an effective PSF that, by
match filtering\footnote{Also called cross-correlation of the images with its PSF. See e.g., Zackay \& Ofek (2015a) for a derivation of the matched filter solution.},
reproduces the optimal transient detection statistic.
Using this image, it is possible to detect and filter out particle hits in both the reference image and the new image, separating these artifacts from real transients.
Another potential use of this image is the optimal detection of
photometric variability and astrometric motion of stars, that works in arbitrarily dense environments.

We demonstrate the efficacy of our algorithm 
on simulated and real images that are
part of the Palomar Transient Factory (PTF; Law et al. 2009), data release 2.

The outline of the paper is as follows:
In \S\ref{sec:overview} we review the state of the art image subtraction
methods, while in \S\ref{sec:derivation} we derive
our optimal transient detection and image subtraction algorithm.
In \S\ref{sec:Properties} we discuss the properties of the derived image subtraction statistic.
A step by step summary of the image subtraction process is presented in \S\ref{sec:algo}.
In \S\ref{sec:tests} we present tests on simulated and real data,
while in \S\ref{sec:code} we describe our code which is available online.
In \S\ref{sec:Details} we discuss the implementation details
and we conclude in \S\ref{sec:summary}.

\section{Brief overview and analysis of existing methods for image subtraction}
\label{sec:overview}

Previously suggested solutions for image subtraction can be divided into two variants.
The first, and more popular variant, can be referred to as regularized partial deconvolution.
Solutions we include in this family are \citet{Phillips95,AL98} and \citet{Bramich}.
\citet{Gal-Yam08} suggested a second variant, which we call cross filtering, 
while \cite{Yuan08} advocated for a mix of the two methods.

Denoting the new image by $N$ and its PSF by $P_{N}$, the reference image by $R$ and its PSF by $P_{R}$,
the first approach attempts to find
a convolution kernel $k$ such that:
\begin{align}
N-k\otimes R \cong 0.
\label{eq:IS}
\end{align}
Here $\otimes$ represents convolution.

The first solution for finding the kernel $k$ was
given by \citet{Phillips95}.
They suggested to perform a deconvolution
solution in Fourier space:
\begin{align}
\widehat{k}=\frac{\widehat{P_n}}{\widehat{P_r}} \cong \frac{\widehat{N}}{\widehat{R}},
\label{eq:deconv}
\end{align}
were $\widehat{\quad}$ represents Fourier transform.
However, this solution is numerically unstable
as the deconvolution operation can (and many times does)
involve division by small numbers.
This problem is apparent from Equation~\ref{eq:deconv},
where the denominator might 
approach zero as fast or faster than the numerator.
Given that any measurement process contains noise,
this division operation amplifies the noise in Fourier space,
which in turn generates correlated noise in real space. The extreme cases of this correlated noise are the characteristic ringing and sinusoidal artifacts that deconvolved images suffer from.

\cite{AL98} suggested
a practical way to mitigate the numerical
instability problem.
Representing $k$ as a set of basis functions,
and noting that Equation~\ref{eq:IS}
is linear, they suggested to solve for
$k$ using linear least squares.
\cite{AL98} suggested to
use a set of basis functions which are
linear combinations of Gaussians
multiplied by low degree polynomials.
Later on, \citet{Bramich} suggested to
solve for the values of a pixelized kernel.
All of the above methods can be viewed as regularization of the deconvolution method of \cite{Phillips95} --
i.e., restricting the solutions for the kernel $k$ to finite size and to some set of logical solutions.
Even though the numerical stability of these algorithms is much better
than that of Equation~\ref{eq:deconv}, they still have several problems.
First, the division by zero problem is still
there and it can become especially pronounced
when the new image has a narrower PSF
(including a PSF that is narrower in any single axis) compared to the reference image.
It is interesting to note that these methods are not symmetric to the exchange of
the new and reference image, while the problem is symmetric to this exchange.
Second, although these methods are intuitive,
they lack statistical justification, there is no rigorous proof they cause no information loss, and it is unclear
what further image processing should be
applied. For example, do we need to apply
another matched filter to the subtracted image in order to detect transients?
If so, which filter should we use?
Third, using these methods the resulting pixel noise is correlated
and there is no simple analytic prescription on how to set
a detection threshold for transient search\footnote{One method to estimate the noise level is using
Bootstrap simulations (e.g., \citealp{Ofeketal2014}).}.
Therefore, it is hard to decide if a detected source is real or an artifact,
or to quantify the probability of it being a false positive.
In addition, in the effort of suppressing the deconvolution artifacts, these solutions sacrifice the cancellation of the constant-in-time image. This will cause large and pronounced subtraction artifacts, that will prevent identification of transients that are substantially fainter than their hosting environment.
Finally, using inversion methods for image subtraction
(i.e., linear least squares)
makes the subtraction process slow, compared with e.g., the Fourier
space solution of \cite{Phillips95}.

The cross filtering solution suggested by \citet{Gal-Yam08}
is to convolve the new image with the PSF of the reference image
and to convolve the reference image with the PSF of the new image:
\begin{align}
S_{\rm GY08} = P_r\otimes N - P_n\otimes R\,.
\end{align}
This solution is always numerically stable, and leaves no subtraction artifacts.
The problem with this solution is, again, the lack of statistical justification,
and that the matched filter for source detection is not specified.

\cite{Yuan08} suggested to apply kernels for both $R$ and $N$,
both chosen from a family of PSFs determined by few parameters,
and to drive the solution towards spatially small kernels by adding the effective PSF area to the loss function.

It is worthwhile to note that the problem of subtracting two images,
and minimizing the resulting difference image in the least square sense
has an infinite number of solutions (see also \citealp{Yuan08}).
For example, the linear equation:
\begin{equation}
K_{{\rm r}} \otimes R - K_{{\rm n}} \otimes N \cong 0,
\label{eq:linDiff}
\end{equation}
where $K_{{\rm r}}$ and $K_{{\rm n}}$ are arbitrary kernels,
has an infinite number of solutions.
This is because for any $K_{{\rm r}}$, we can find
$K_{{\rm n}}$ that satisfies Equation~\ref{eq:linDiff} in
the least squares sense.
It is clear from this simple analysis that
all subtraction methods mentioned are focused on making
the PSF of the two images identical, with very little
attention to the maximization of the signal-to-noise ratio ($S/N$)
of a transient source that appears in one of the images.
In a sense, these methods do~not solve the transient detection problem,
but a different problem which is how to make two images as similar as possible using convolution.
In this paper, we rigorously derive a method that cancels the constant-in-time image {\it and}
maximizes the $S/N$ of a transient source at the same time.
We note that there are several ways to derive this method.
Here we will derive it from first principles via modeling
the transient detection with simple hypothesis testing and using the lemma of Neyman \& Pearson (1933).



\section{Statistical Derivation}
\label{sec:derivation}

Given the numerous problems with existing
image subtraction methods, we would like to place the transient
detection problem on firm statistical grounds.
In \S\ref{subsec:NewRefSubtraction} we outline the derivation and formulae
of our image subtraction statistics.
Given that the full derivation is tedious we defer it to Appendix~\ref{App:FullDeriv}.
In \S\ref{subsec:ImageSubWithManyRef} we show that the best way to build
a reference image, for the purpose of image subtraction,
is to use the image coaddition algorithm of Zackay \& Ofek (2015b).
Our derivation in \S\ref{subsec:NewRefSubtraction} assumes
that the images are background-noise dominated
(i.e., the objects we care about
have source noise which is lower than the background noise).
This causes an underestimation of the noise near bright sources.
In \S\ref{sec:SourceNoise} we present a simple correction to the image subtraction
formulae that takes care of the source noise and other errors,
like registration noise.
In \S\ref{sec:AstNoise} we present an accurate treatment of astrometric
shifts, noise and color-refraction errors.
In \S\ref{sec:Gain} we outline our suggested method to equalize
the flux zero points of the new and reference images.
In \S\ref{sec:PSFphot} we provide an algorithm for optimal PSF photometry
in the subtraction image,
while in \S\ref{sec:CR} we describe how this method can be used
for cosmic-ray, bad pixels and reflection-ghost identification.

\subsection{Transient source detection using image subtraction}
\label{subsec:NewRefSubtraction}

Here we derive, from first principles, an optimal method for transient source
detection, under the assumptions that the images are
background-noise dominated,
and the noise is Gaussian and independent\footnote{In practice the pixels maybe slightly correlated due to charge repulsion and charge diffusion in a CCD.}.

Let $R$ and $N$ be the background-subtracted reference image and the
background-subtracted new image, respectively.
Denote by $T$ the background-subtracted true constant sky image.
Denote by $P_r$ and $P_n$ the point spread functions (PSFs)  of the reference image and the new image, respectively.
$P_r$ and $P_n$ are normalized to have unit sum.
We assume that $P_{n}$, $P_{r}$, and the flux-based zero
points\footnote{Following Zackay \& Ofek (2015a, 2015b)
this factor represents the product of atmospheric transparency, telescope and detector transmission
and integration time.}
of the new image ($F_{n}$) and reference image ($F_{r}$) are known.
We present a method for finding $F_{n}$ and $F_{r}$ in \S\ref{sec:Gain},
and the PSF measurements are discussed in \S\ref{sec:PSF}.

The expression for the reference image is:
\begin{align}
R = F_{r} T\otimes P_r + \epsilon_r,
\label{eq:Rdef}
\end{align}
where $\epsilon_r$ is the additive noise component of the image $R$.

Given the null hypothesis,  $\mathcal{H}_0$, that states there are no new sources in the new image we can write:
\begin{align}
N_{|\mathcal{H}_0} = F_{n} T\otimes P_n + \epsilon_n.
\label{eq:NdefH0}
\end{align}
Given the alternative hypothesis, $\mathcal{H}_1(q,\alpha)$,
that states there is a new
point source at position $q$ with flux $\alpha$ in the new image, we can write:
\begin{align}
N_{|\mathcal{H}_1(q,\alpha)} = F_{n} T\otimes P_n + \alpha F_{n} \delta(q)\otimes P_n  + \epsilon_n,
\label{eq:NdefH1}
\end{align}
where
$\delta(q)$ denotes a two dimensional image with one at position $q$, and zero otherwise.
We assume that the dominant source of noise is the background noise,
$\epsilon_r$ and $\epsilon_n$ both satisfy that all pairs of pixels are uncorrelated --
i.e., that for all pairs of pixels $x_1,x_2$ for which $x_1\neq x_2$:
\begin{align}
{\rm Cov}\left(\epsilon_r[x_1],\epsilon_r[x_2]\right) = 0 \,,{\rm Cov}\left(\epsilon_n[x_1],\epsilon_n[x_2]\right) = 0,
\end{align}
and that all pixels have spatially uniform variance\footnote{As the convolution is a local operation, this assumption can be relaxed (see discussion in Zackay \& Ofek 2015a).}:
\begin{align}
V(\epsilon_r[x]) = \sigma_r^2\,,V(\epsilon_n[x]) = \sigma_n^2.
\end{align}

Because both hypotheses are simple\footnote{A simple hypothesis has no unknown parameters.
We are applying the hypothesis testing to each value of $\alpha$ and $q$ separately.},
we can use the Neyman-Pearson lemma \citep{NeymanPearsonLemma},
that states that the most powerful\footnote{The power of a binary hypothesis test is the probability that the test correctly rejects the null hypothesis when the alternative hypothesis is true.}
statistic for deciding between two simple hypotheses is the likelihood ratio test:
\begin{align}
\mathcal{L}(q,\alpha) = \frac{\mathcal{P}(N,R|\mathcal{H}_0)}{\mathcal{P}(N,R|\mathcal{H}_1(q,\alpha))},
\end{align}
where $\mathcal{P}$ denotes probability.
A critical point is that we do not have any prior information or assumptions on $T$. Therefore, we cannot calculate the probabilities  $\mathcal{P}(N,R|\mathcal{H}_0)$ and $\mathcal{P}(N,R|\mathcal{H}_1(q,\alpha))$ directly.
However, we can calculate their ratio by developing the expression using the law of conditional probabilities
\begin{align}
\mathcal{L}(q,\alpha) = \frac{\mathcal{P}(N|R,\mathcal{H}_0)\mathcal{P}(R|\mathcal{H}_0)}{\mathcal{P}(N|R,\mathcal{H}_1(q,\alpha))\mathcal{P}(R|\mathcal{H}_1(q,\alpha))}.
\end{align}
Next we can use the fact that $\mathcal{H}_0$ and $\mathcal{H}_1$
predict the same likelihood to the reference and cancel
out the last multiplicative terms in the numerator
and denominator.

After some algebra, which is detailed in Appendix~\ref{App:FullDeriv},
we can find the optimal statistic for source detection
\begin{align}
\widehat{S} \equiv \widehat{\frac{1}{\alpha}\log{\mathcal{L}}} = \frac{F_{n}F_{r}^{2}\overline{\widehat{P_n}}|\widehat{P_r}|^2\widehat{N} - F_{r}F_{n}^{2}\overline{\widehat{P_r}}|\widehat{P_n}|^2\widehat{R} }{\sigma_r^2F_{n}^{2}|\widehat{P_n}|^2 + \sigma_n^2F_{r}^{2}|\widehat{P_r}|^2},
\label{eq:S}
\end{align}
where the over-line symbol denotes the complex conjugate operation.
We note that by putting the over-line sign above the hat sign we mean that the
complex conjugate operation follows the Fourier transform operation.
%
%
This statistic (or score image) is simply the log-likelihood ratio test
between the two hypotheses.
This score is calculated simultanously for all values of $\alpha$,
while each pixel in the score image refers to a different $q$ position.
It is important to note that Equation~\ref{eq:S} is a matched filter
image and no further filtering is required.
In order to find transients all we need to do 
is to identify local maxima (or minima) in $S$.
The significance of a local maximum, in units of sigmas, is given by its value
divided by the standard deviation of the image $S$.

Since Equation~\ref{eq:S} is a matched filter image, its pixels
are correlated, and any hypothesis testing or measurement, other than
transient detection and photometry (see \S\ref{sec:PSFphot}),
requires a knowledge
of the covariance between the pixels.
An example for such hypothesis testing is cosmic-ray identification
via image subtraction, or searching for variable nebulosity (e.g., light echos).
In order to have an image-subtraction method that is optimal for all purposes
and easy to use,
we need to identify an image whose pixel noise is uncorrelated,
and that cross-correlating this image with its own PSF returns Equation~\ref{eq:S}.
In Appendix~\ref{App:FullDeriv} we identify such an image as:
\begin{align}
\widehat{D} = \frac{{F_{r}\widehat{P_r}}{\widehat{N}} - {F_{n}\widehat{P_n}}{\widehat{R}}}{\sqrt{\sigma_n^2F_{r}^{2}|\widehat{P_r}|^2 + \sigma_r^2F_{n}^{2}|\widehat{P_n}|^2}}\,.
\label{eq:D}
\end{align}
The PSF of this image, normalized to have unit sum, is given by:
\begin{align}
\widehat{P_D} = \frac{F_rF_n\widehat{P_r}\widehat{P_n}}{F_D\sqrt{\sigma_n^2F_{r}^{2}|\widehat{P_r}|^2 + \sigma_r^2F_{n}^{2}|\widehat{P_n}|^2}},
\label{eq:Dpsf}
\end{align}
where $F_D$ is the flux-based zero point of the subtraction image, which is given by:
\begin{align}
F_{D} = \frac{F_{r}F_{n}}{\sqrt{\sigma_{n}^{2}F_{r}^{2} + \sigma_{r}^{2}F_{n}^{2}}}
\label{eq:Fd}
\end{align}

Indeed, using this difference image $D$ and its PSF
we can verify that the cross-correlation of $D$ with $P_{D}$ returns:
\begin{align}
S = F_{D} D\otimes\overleftarrow{P_D},
\label{eq:SDP}
\end{align}
where the backward arrow sign denotes coordinate reversal
(i.e., $\overleftarrow{P}(x,y)=P(-x,-y)$).
Alternatively in Fourier space
\begin{align}
\widehat{S} = F_{D}\widehat{D}\overline{\widehat{P_{D}}}.
\label{eq:Sfs}
\end{align}

It is important to note that,
in the background-dominated noise limit, $D$ is a proper image, and
hence we call it {\it the proper subtraction image}.
As in Zackay \& Ofek (2015b) we define a proper image to
be an image whose noise is independent and identically\footnote{In practice the noise levels need to be identical
only locally (on scales which are twice the PSF size), as the convolution is a local operation. In the vicinity of bright stars $D$ is not proper.}
distributed (i.i.d).
This means that $D$ can be used for any hypothesis testing or measurement,
without the need for the covariance between the pixels.
Furthermore, in Appendix~\ref{Ap:Sufficiency}
we present a proof that $D$ and $P_{D}$ are in fact
sufficient statistics\footnote{In statistics, a statistic is sufficient with respect to a statistical model and its associated unknown parameter if no other statistic that can be calculated from the same sample provides any additional information as to the value of the parameter.}
for any hypothesis testing or measurement.

Equation~\ref{eq:D} and its PSF (Eq.~\ref{eq:Dpsf}) are adequate for detection
of objects whose original shape was convolved with the telescope and atmosphere PSF.
However, particle hit events do~not share this PSF.
In Appendix~\ref{Ap:Sufficiency} we derive the PSF in the difference image $D$, of a $\delta$-function in $N$ or $R$.
The PSF in the difference image $D$ of a $\delta$-function in $N$ is:
\begin{equation}
\widehat{P_{D_N}} = \frac{F_{r}\widehat{P_r}}{F_{D_n}\sqrt{\sigma_n^2F_{r}^{2}|\widehat{P_r}|^2 + \sigma_r^2F_{n}^{2}|\widehat{P_n}|^2}}\,,
\label{eq:Pdn}
\end{equation}
while the PSF in the difference image $D$ of a $\delta$-function in $R$ is:
\begin{equation}
\widehat{P_{D_R}} = \frac{F_{n}\widehat{P_n}}{F_{D_R}\sqrt{\sigma_n^2F_{r}^{2}|\widehat{P_r}|^2 + \sigma_r^2F_{n}^{2}|\widehat{P_n}|^2}}\,.
\label{eq:Pdr}
\end{equation}
These PSFs are also accompanied by the corresponding zero-points, $F_{D_N},F_{D_R}$ that can be found in Appendix~\ref{Ap:Sufficiency}.

These equations are useful if one would like to search
for events which are similar to
a delta function (e.g., bad pixels).
We note that $\widehat{P_{D_N}}$ and $\widehat{P_{D_R}}$ in many cases
can be approximated by a delta function.

To summarize, in order to find a transient source in either the
reference or the new image we can calculate $D$ (Eq.~\ref{eq:D})
and cross-correlate it with its PSF (Eq.~\ref{eq:Dpsf}).
Alternatively, we can calculate directly the statistic $S$ (Eq.~\ref{eq:S}).

\subsection{Construction of the reference image}
\label{subsec:ImageSubWithManyRef}

Typically, the reference image is built by coadding multiple images.
Here we will show that the best way to produce
a reference image for subtraction is using the method
described in Zackay \& Ofek (2015b).

In the case of multiple reference images
we need to replace Equation~\ref{eq:Rdef}
with the model for the $j$-th reference image:
\begin{equation}
R_j = F_{j}P_{j}\otimes T + \epsilon_j.
\label{eq:Rmdef}
\end{equation}
Here $F_{j}$ is the flux-based zero point of the $j$-th reference image,
$P_{j}$ is the PSF of the $j$-th reference image,
and $\epsilon_{j}$ is the noise of the $j$-th reference image.

As before, the model for $N$ assuming the null hypothesis, $\mathcal{H}_0$, is
given by Equation~\ref{eq:NdefH0},
while if the first hypothesis,  $\mathcal{H}_1$, is true
then $N$ is given by Equation~\ref{eq:NdefH1}.

As in the previous section, we would like
to decide between two simple hypotheses.
Therefore, the optimal test statistic is the likelihood ratio test \citep{NeymanPearsonLemma}
\begin{align}
\mathcal{L}(q,\alpha) = \frac{\mathcal{P}(N,R_1,\dots,R_J|\mathcal{H}_0)}{\mathcal{P}(N,R_1,\dots,R_J|\mathcal{H}_1(q,\alpha))}\,.
\end{align}
As before, we can use the law of conditional probabilities, and the fact that $\mathcal{H}_0$ and $\mathcal{H}_1$
predict the same likelihood for all references.
The full derivation is presented in Appendix~\ref{secA:RefIm}, and after some algebra
we find that the optimal reference image is given by
\begin{align}
\widehat{R} = \frac{\sum_j\frac{F_{j}}{\sigma_j^2} \overline{\widehat{P}_j}  \widehat{R_j}}{\sqrt{\sum_j\frac{F_{j}^{2}}{\sigma_j^2}|\widehat{P}_j|^2 }}.
\label{eq:ProperCoadd}
\end{align}
%
The PSF (normalized to have unit sum) of the reference image is given by:
\begin{align}
\widehat{P_{R}} = \frac{\sqrt{\sum_{j}{\frac{F_{j}^{2}}{\sigma_{j}^{2}}|\widehat{P_{j} }|^{2} } } }{F_{r}},
\label{eq:ProperCoaddPSF}
\end{align}
where $F_{r}$ is the flux-based zero point of the reference:
\begin{align}
F_{r} = \sqrt{\sum_{j}{\frac{F_{j}^{2}}{\sigma_{j}^{2}}  } }.
\label{eq:Fr}
\end{align}

Not surprising, this is identical to the optimal coaddition method
derived in Zackay \& Ofek (2015b).
We note that the reason $R$ preserves all the information
from the individual references
is because in the computation of each frequency in $R$,
we add random variables scaled by their (conjugate) expectation,
divided by the variance. We can identify this operation
as the maximal $S/N$ addition of random variables
(see Appendix~A of Zackay \& Ofek 2015a).
The reader should refer to Zackay \& Ofek (2015b) for analysis and proof of sufficiency of this
so called {\it proper coaddition} method.

\subsection{Simple, suboptimal correction for source noise, astrometric noise and color-refraction noise}
\label{sec:SourceNoise}


Equation~\ref{eq:S} ignores the source noise, and hence
the noise level is underestimated in the vicinity of bright stars.
The outcome of this will be that bright sources may be flagged
as possible transients or variables.
Furthermore, this equation ignores any additional important sources 
of noise like astrometric noise, astrometric scintillation noise,
color-refraction noise,
flux scintillation noise,
and position-dependent flat-fielding errors.

A simple correction to this problem, albeit suboptimal, is to divide $S$ by
a correction factor that takes into account the local
estimated variance of the extra noise.
Derivation of this correction factor is presented in Appendix~\ref{Ap:CorrectSourceNoise}.
In the image space, the expression for the corrected $S$ is
\begin{equation}
S_{{\rm corr}} = \frac{S}{\sqrt{V(S_{N}) + V(S_{R}) + V_{{\rm ast}}(S_{N}) + V_{{\rm ast}}(S_{R}) + ...  }}.
\label{eq:Scorr}
\end{equation}
Here the terms in the denominator may include any position-dependent contribution to the variance,
that is not included in the $\sigma_{n}^{2}$ and $\sigma_{r}^{2}$ factors.

In this example we list two specific contributions from the source noise
and from astrometric noise.
The first two terms in the denominator are the variance from the source
noise in the new and reference images, respectively,
while the next two terms are the variance due to astrometric noise.
Other sources of noise like color-refraction can be added in a similar manner.

Here $V(S_{N})$ is the variance of the part of $S$ containing $N$ given by
\begin{equation}
V(S_{N}) = V(\epsilon_{n}) \otimes (k_{n}^{2}),
\label{eq:VSN}
\end{equation}
and $V(S_{R})$ is the variance of the part of $S$ containing $R$ given by
\begin{equation}
V(S_{R}) = V(\epsilon_{r}) \otimes (k_{r}^{2}),
\label{eq:VSR}
\end{equation}
and the Fourier transform of $k_{r}$ is given by
\begin{equation}
\widehat{k_{r}} = \frac{F_r F_n^{2}\overline{\widehat{P_r}}|\widehat{P_n}|^2 }{\sigma_r^2F_n^2|\widehat{P_n}|^2 + \sigma_n^2F_r^2|\widehat{P_r}|^2},
\label{eq:kr}
\end{equation}
while the Fourier transform of $k_{n}$ is
\begin{equation}
\widehat{k_{n}} = \frac{F_n F_r^{2}\overline{\widehat{P_n}}|\widehat{P_r}|^2 }{\sigma_r^2F_n^2|\widehat{P_n}|^2 + \sigma_n^2F_r^2|\widehat{P_r}|^2}.
\label{eq:kn}
\end{equation}
The variance of $\epsilon_{n}$ and $\epsilon_{r}$ are simply
the variance images.
For a single image the variance map, $V(\epsilon_{n})$, is simply the number of electrons
in each pixel (including the background),
added with the readout noise squared.
However, in the case of multiple images, the correct way to construct $V(S_{R})$ is
to calculate $k_{r}$, $V(\epsilon_{r})$, and $V(S_{R})$ for each reference
image and to sum all the individual $V(S_{R})$ values up (see Appendix~\ref{secA:RefIm}).
However, in many cases a reasonable approximation is to calculate $k_{r}$
from the properly coadded image, and  calculate $V(\epsilon_{r})$
using a simple addition of all the images (in units of electrons) from which the reference was constructed
(i.e., the number of electrons in each pixel including the background)
added with the total readnoise squared.

Next, the astrometric variance terms are given by
\begin{align}
V_{{\rm ast}}(S_{N}) = \sigma_{x}^{2}\Big(\frac{dS_{N}}{dx}\Big)^{2}+\sigma_{y}^{2}\Big(\frac{dS_{N}}{dy}\Big)^{2},
\label{eq:VastSN}
\end{align}
where $\sigma_{x}$ and $\sigma_{y}$ are the astrometric registration noise
in the $x$ and $y$ axes, respectively,
while $\frac{dS_{N}}{dx}$ and $\frac{dS_{N}}{dy}$ are the gradients of $S_{N}$ in the $x$ and $y$ directions.
Here the Fourier transform of $S_{N}$ is given by
\begin{align}
\widehat{S_{N}} = \widehat{k_{n}}\widehat{N}.
\label{eq:SN}
\end{align}
In a similar manner
\begin{align}
V_{{\rm ast}}(S_{R}) = \sigma_{x}^{2}\Big(\frac{dS_{R}}{dx}\Big)^{2}+\sigma_{y}^{2}\Big(\frac{dS_{R}}{dy}\Big)^{2}.
\label{eq:VastSR}
\end{align}
Here the Fourier transform of $S_{R}$ is given by
\begin{align}
\widehat{S_{R}} = \widehat{k_{r}}\widehat{R}.
\label{eq:SR}
\end{align}
The origin of these terms is that astrometric noise causes shifts
in individual PSFs. The noise induced by these shifts is proportional
to the difference between neighboring pixels (i.e., the gradient).

We note that in practice the astrometric registration noise is
the rms of the registration fitting process.
This term include both registration errors and the astrometric scintillation noise.
In some cases the quality of the registration is position dependent.
In this case it is possible to replace the scalars $\sigma_{X}$ and $\sigma_{Y}$
by matrices of the position-dependent noise.
In \S\ref{sec:AstNoise} we suggest a more accurate treatment of the astrometric noise component.

\subsection{Accurate treatment of astrometric noise and flux variability}
\label{sec:AstNoise}

Astrometric errors and shifts are a major problem for image subtraction.
For example, for a bright source with $10^{4}$\,electrons
and full-width at half maximum (FWHM) of 2\,pixels,
the astrometric error induced by the Poisson noise
will be about a few tens of milli-pixels.
This is equivalent to the 
typical astrometric scintillation noise
induced by the Earth turbulent atmosphere
(see \S\ref{sec:scint}).
Therefore, even in the case of high quality registration,
we expect that all bright stars will have
subtraction residuals due to astrometric scintillation noise.

Fortunately, due to the closed form and numerical stability of our method,
the shape of the subtraction residuals
is fully predictable, given the astrometric shift
and the flux difference between the star
as it appears in the reference and as it appears in the new image.
Therefore, we can use this to measure
the astrometric shift and flux variability for each star.

For adequately\footnote{By adequately sampled images
we mean that the PSF width is sampled by at least two pixels.
This can be referred to as the Nyquist sampling of the PSF
by the camera.}
sampled images, this proposed mechanism is accurate,
and it allows us to measure astrometric shifts and variability
in very crowded fields.
%
The details of this method will be presented in a future publication,
but here we provide a brief outline:
The astrometric shift and photometric variability kernel is:
%
\begin{align}
\widehat{P}_{S}(\alpha_n,\alpha_r,\Delta x, \Delta y) = \widehat{P_D}\left({\alpha_r} - {\alpha_n} \widehat{s}\right).
\label{eq:shift}
\end{align}
%
%
Here $\alpha_n$ is the flux of the source
in $N$ and $\alpha_r$ is its flux in $R$,
and $\widehat{s}$ is the shift operator
(including sub-pixel shifts) in Fourier space.
This operator is a function of the shifts $\Delta{x}$
and $\Delta{y}$.
%
%
%
%
%
Using Equation~\ref{eq:shift}, 
we can treat residuals detected in $S$ more
carefully than we did in \S\ref{sec:SourceNoise}.
Specifically, we can now perform hypothesis testing to
decide between e.g., $\mathcal{H}_0$: changes are consistent with
stationary and non variable source;
or $\mathcal{H}_1$: the star moved or its flux changed.
This scheme can be applied to any part of $D$,
for which we identify a significant peak in $S$ (e.g., above 3$\sigma$).
Apart from using this to eliminate false positives,
we can now use this to detect and measure new kinds of signals.
For example, we can use it to search for moving objects blindly, even in the presence of complex, constant in time, structure in the background.

%
%
%



\subsection{Matching the local zero points, background flux and astrometric shift}
\label{sec:Gain}

Our solution so far assumed that the values
of the flux-based zero points ($F_{r}$ and $F_{n}$),
the background levels, ($B_{n}$ and $B_{r}$),
and the relative astrometric shift ($\Delta{x}$ and $\Delta{y}$)
are known.
Careful analysis of Equation~\ref{eq:D}
shows that, in practice, we only care about
the flux zero points ratio
\begin{align}
\beta \equiv F_{n}/F_{r},
\label{eq:beta}
\end{align}
the background difference,
\begin{align}
\gamma \equiv B_{n} - B_{r},
\label{eq:gamma}
\end{align}
and the translation ($\Delta{x}$, $\Delta{y}$).

By substituting Equations~\ref{eq:beta} and \ref{eq:gamma} into $D$ (Eq.~\ref{eq:D}),
and introducing the shift operator we can get the desired expression
we need to minimize in order to find $\beta$, $\gamma$, $\Delta{x}$, and $\Delta{y}$.
This can be done either  locally (in small sections of the image),
or globally.
For simplicity, and since we already discussed astrometric shifts
in \S\ref{sec:AstNoise}, here we neglect translations,
but toward the end we will mention how this can be incorporated.

In order to find $\beta$ and $\gamma$ we need to compare
the two parts of $\widehat{D}$:
\begin{align}
\widehat{D_{n}}(\beta) = \frac{\widehat{P_{r}}\widehat{N}}{\sqrt{\sigma_{n}^{2}|\widehat{P_{r}}|^{2} + \beta^{2}\sigma_{r}^{2}|\widehat{P_{n}}|^{2} }},
\label{eq:Dn}
\end{align}
and
\begin{align}
\widehat{D_{r}}(\beta) = \frac{\widehat{P_{n}}\widehat{R}}{\sqrt{\sigma_{n}^{2}|\widehat{P_{r}}|^{2} + \beta^{2}\sigma_{r}^{2}|\widehat{P_{n}}|^{2} }}.
\label{eq:Dr}
\end{align}
Note that we replaced $F_{n}$ and $F_{r}$ by $\beta$.
All we need to do is to inverse Fourier transform
$\widehat{D_{n}}$ and $\widehat{D_{r}}$ and to solve the following
non-linear equation for $\beta$ and $\gamma'$ (and optionally
$\Delta{x}$ and $\Delta{y}$):
\begin{align}
D_{n}(\beta) = \beta D_{r}(\beta) + \gamma'
\label{eq:DnDrBeta}
\end{align}
where
\begin{align}
\gamma' = \frac{\gamma}{\sqrt{\sigma_{n}^{2} + \beta^{2}\sigma_{r}^{2} }}.
\label{eq:gammatag}
\end{align}
Note that the solution should be performed in the image domain.
If we are interested in solving also for small translations,
we need to multiply $\widehat{D_{r}}$ and $\gamma'$
with the shift operator.
If we trust that the images were background subtracted and aligned correctly,
then we can set $\gamma=0$, $\Delta{x}=0$, $\Delta{y}=0$ and
use the same expression to solve only for the value of $\beta$.

Equation~\ref{eq:DnDrBeta} is non-linear in $\beta$.
Therefore iterative solutions are required.
For example,
in the first iteration set $\beta=1$ and solve for the new value of $\beta$,
and use it in the next iteration to find a new value of $\beta$,
until convergence\footnote{We found that usually $\beta$ converges in
2--3 iterations.}.
Furthermore, it is important to note that one must use robust fitting
methods in order to solve Equation~\ref{eq:DnDrBeta}.
The reason is that there may be bad pixels,
particle hits, astrometric noise, and saturated pixels in the images.
It is also recomended to remove the images-edge pixels prior to fitting $\beta$.

\subsection{PSF photometry in the difference image}
\label{sec:PSFphot}

In this section we present a statistic for measuring
the PSF photometry\footnote{PSF photometry refers to (effectively) fitting the source
with a PSF.}
of a source in the difference image.
This measurement statistic is unbiased and has
maximal $S/N$ among all estimators which are linear
combinations of the input images.
However, this statistic is optimal only for the
background-dominated-noise limit.
A full derivation of this statistic is presented in Appendix~\ref{Ap:optimalPhotometry}.

The best linear estimator for the PSF photometry of a source at position $q$
is
\begin{equation}
\widetilde{\alpha(q)}=\frac{S(q)} {F_{S}}.
\label{eq:FluxEst}
\end{equation}
Here $F_{S}$ is the flux normalization of $S$:
\begin{align}
F_{S} =  \sum_{f}{\frac{F_{n}^{2}|\widehat{P_n}|^2 F_{r}^{2}|\widehat{P_r}|^2}{\sigma_r^2F_{n}^{2}|\widehat{P_n}|^2 + \sigma_n^{2}F_{r}^{2}|\widehat{P_r}|^2}},
\label{eq:F_S}
\end{align}
where $f$ indicates spatial frequencies.
The standard deviation of this estimator is
\begin{equation}
\sigma_{\widetilde{\alpha(q)}} = \frac{\sqrt{V(S_N) +{V(S_{R})}}}{F_{S}},
\label{eq:FluxStdEst}
\end{equation}
where $V(S_N)$ and $V(S_R)$ are defined in Equations~\ref{eq:VSN}--\ref{eq:VSR}.
Note that Equation~\ref{eq:FluxEst} can be used to measure
the PSF flux of all the transients in the image simultanously.

\subsection{Cosmic ray, bad pixel and ghosts identification}
\label{sec:CR}

The image subtraction statistic $D$ can be used to identify
cosmic rays and bad pixels.
A major advantage of using the proper image subtraction
over other image-differencing techniques
is that its pixels noise is uncorrelated and usually
it roughly preserves the shape of sources which are similar to $\delta$-functions.
This means that in most cases one can
identify particle hits by applying edge-detection algorithms
(e.g., \citealp{vanDokkum2011}),
without any modifications, directly on $D$.

An alternative approach is to use a rough model for the shapes of particle hits and bad pixels,
and to perform a composite hypothesis testing.
The log-likelihood of observing $D$, if
an object at position $q$ is a point source transient
($\mathcal{H}_{\rm ps}$ hypothesis) with flux $\alpha$, is given by
\begin{align}
-\log(\mathcal{P}(D|\mathcal{H}_{\rm ps}(q)))=\sum_x{||D - \alpha F_D\overleftarrow{P_{D}}\otimes\delta(q)||^2},
\label{eq:logPps}
\end{align}
while the log-likelihood of $D$ if the object
at position $q$ is a cosmic ray with flux $\alpha$
and with shape $P_{{\rm cr}}$ in $N$,
($\mathcal{H}_{\rm cr}$ hypothesis) is
\begin{align}
-\log(\mathcal{P}(D|\mathcal{H}_{\rm cr}(q)))=\sum_x{||D-\alpha P_{{\rm cr}}\otimes \overleftarrow{P_{D_N}}\otimes \delta(q)||^2}.
\label{eq:logPcr}
\end{align}
Here $x$ is the subset of pixels that contain the source of interest
(e.g., an area with a width twice that of the PSF around the source).
The difference between Equations~\ref{eq:logPps} and \ref{eq:logPcr}
(using appropriate priors, such as the probability of seeing a transient at a certain magnitude and the probability of seeing a cosmic ray with this flux)
is a statistic that can be indicative
(after setting the appropriate threshold)
for deciding whether the detected transient
is a cosmic ray or an astronomical transient.
We note that in this case the flux of the source, and the intensity and shape of the cosmic ray are free parameters of the model.
Therefore, this is a classic case of composite hypothesis testing.

The same approach can be used to identify
internal-reflection ghosts.
In this case we need to replace the shape $P_{{\rm cr}}$ with the
shape of a reflection ghost. For example, an extended kernel
(e.g., top hat filter)
which is wider than the stellar PSF.


\section{Properties of the new image subtraction method}
\label{sec:Properties}

Now that we have an optimal solution for the subtraction problem,
we can analyze its properties and compare it to other methods, seeking an intuitive understanding.

\subsection{Optimality}

Our image subtraction and transient detection formulae were derived using
the lemma of Neyman \& Pearson (1933).
This ensures that whenever our assumptions are correct
our method is optimal.
Our assumptions:
the images are registered, dominated by uncorrelated Gaussian background noise,
and that the PSFs, background, variance and flux-based zero points are known.

\subsection{The constant-in-time image $T$ cancels}

For perfectly registered images,
both the optimal proper difference image ($D$)
and transient detection image ($S$) are free
of subtraction residuals from the constant-in-time image.
This is because the constant-in-time image $T$ algebraically vanishes.

This is not the case in the subtraction methods suggested by \cite{AL98} and \cite{Bramich}.
In these methods an optimum for the trade-off between
magnifying the image noise and minimizing the
constant-in-time residuals of $T$ was explored.

\subsection{Numerical stability}
\label{sec:NumStab}

Inspecting Equations \ref{eq:S}, \ref{eq:D} and \ref{eq:Dpsf},
it is apparent that if the denominator is approaching zero,
then the numerator is approaching zero even faster.
Therefore our image subtraction method is numerically stable for all combinations of PSFs for the new and reference images.


We note that the \cite{AL98} and \cite{Bramich}
methods are numerically unstable in
the general case, as these methods effectively perform deconvolution.
It is true that if the PSF of the reference image is narrower, in all axes,
than the PSF of the new image,
then the \cite{AL98} family of methods are stable.
However, even in this case the solution found by these methods
is sub-optimal (i.e., it does not maximize the $S/N$ of the transients).
We further demonstrate this point in \S\ref{sec:tests}.

\subsection{Locality}

An important property of our image-subtraction statistic is its locality with respect to the input data.
The formulae for the image subtraction statistic are stated in Fourier domain for simplicity and clarity.
But, even though operations done in the Fourier domain are global in nature, when calculating the
final kernels that
actually multiply $\widehat{R}$ and $\widehat{N}$,
we see that their representation in the spatial domain is local,
with power vanishing quickly away from the origin (i.e., the PSFs are approaching zero at
large distance from their origin).
This is because these operations represent convolution with a finite-size kernel.
Therefore, the proper subtraction statistic could be calculated independently for every arbitrarily small image patch
(up to few times the PSF size), allowing the PSF to vary smoothly across the image.
In addition, local artifacts such as bad pixels, particle hits or saturated stars will affect only their close vicinity
due to the locality of the kernels used.

\subsection{The proper image subtraction $D$ has white noise}
\label{sec:whitenoise}

In the expression for $\widehat{D}$ (Eq. \ref{eq:D}),
in the background-noise dominated limit,
the variance of the numerator is equal to the square of the denominator, i.e:
\begin{align}
V[F_r\widehat{P_r}\widehat{N} - F_n\widehat{P_n}\widehat{R}] = \sigma_r^2F_n^2|\widehat{P_n}|^2 + \sigma_n^2F_r^2|\widehat{P_r}|^2 ,
\end{align}
which means that all the spatial frequencies of $\widehat{D}$ have equal variance. 
Furthermore, since we assume that the images have white noise, their Fourier transform has white noise.
This means that the spatial frequencies of $\widehat{D}$, as a linear combination of $\widehat{R},\widehat{N}$, has un-correlated noise. Together, both properties mean that $\widehat{D}$ has white noise, which means that $D$ has also white noise. In other words, the difference image is
a proper image (as defined in Zackay \& Ofek 2015b).
This property is violated by all the other methods
for image subtraction.


We note that in the vicinity of bright stars, where the source noise variance is dominant,
the proper subtraction image $D$ exhibits correlated noise.
Our simulations suggest that if the source variance is at least
an order of magnitude higher than the
background variance, than correlated noise is detectable by eye
in the vicinity of such sources.
However, as we stated before, using our method,
the source noise is controllable via variance corrections.

\subsection{$D$ and $P_{D}$ are sufficient for any measurement or decision on the difference between the images}

In statistics, a statistic is sufficient with respect to a statistical model and its associated unknown parameters if
no other statistic that can be calculated from the same sample provides any additional information as to the value of the parameter.
In Appendix~\ref{Ap:Sufficiency} we provide a proof that $D$ and $P_{D}$ are sufficient for any measurement or
hypothesis testing on the difference between the images.
The key ingredients for this proof are that any likelihood calculation for any generative model for the difference between the images
can be computed by using only these quantities, and the use of the
Fisher-Neyman factorization theorem (Fisher 1922; Neyman 1935).
We note that there are infinite number 
of sufficient statistics with respect to the
image subtraction problem
(see some examples in Zackay \& Ofek 2015b
in the context of coaddition).
Here we prefer the proper subtraction image $D$ (rather than e.g., $S$) due to its useful properties.

The sufficiency property has important practical consequences.
It means that, in the background-noise dominated limit,
$D$ and $P_{D}$ contain all the information one needs for any further
measurement or hypothesis testing related to the difference between the images.
There is no need for other types of difference images for other applications.
Examples for practical applications include the identification and removal of particle hits
on the detector (see \S\ref{sec:CR});
optimal search for proper motion, astrometric shifts (\S\ref{sec:AstNoise}) and asteroid streaks.

\subsection{Symmetry between the new image and the reference image} 

The problem of image subtraction is
symmetric to the exchange of the reference and the new image (up to negation of the flux of the transient).
Therefore, it is not surprising that the optimal image
subtraction statistics ($D$ or $S$) are symmetric
to the exchange of $R$ and $N$ (up to a minus sign).
This property is violated by the solutions proposed by \citet{Phillips95},
\citet{AL98}, and \citet{Bramich}.
We note that the \cite{Gal-Yam08} method preserves this symmetry.
Interestingly, the \cite{Gal-Yam08} method is identical to the numerator
of the proper image subtraction statistic ($D$).

\subsection{The limit of noiseless reference image}
\label{sec:NoislessR}

In the limit of $\sigma_{r}\rightarrow0$ Equation~\ref{eq:S} becomes
\begin{align}
\lim_{\sigma_{r}\rightarrow0}{\widehat{S}} = \frac{F_nF_r^2\overline{\widehat{P_n}}|\widehat{P_r}|^2\widehat{N} - F_rF_n^2\overline{\widehat{P_r}}|\widehat{P_n}|^2\widehat{R} }{\sigma_n^2F_r^2|\widehat{P_r}|^2} \\
=  \frac{F_n\overline{\widehat{P_n}}}{\sigma_n^2} \big( \widehat{N} - \frac{F_n\widehat{P_{n}}}{F_r\widehat{P_r}} \widehat{R} \Big).
\label{eq:SlimSigmar0}
\end{align}
The term $\widehat{P_{n}}/\widehat{P_{r}}$ can be identified
as the convolution kernel solved for by the methods of
\citet{Phillips95}, \citet{AL98} and \citet{Bramich}.
Therefore, in this limit, $S$ converges to the \citet{AL98} family of methods
followed by filtering each of the images with the PSF of the new image.

This simple analysis demonstrates that the \citet{AL98} family of methods, if followed by the correct
matched filtering, is a special case of our solution $S$.
Furthermore, Equation~\ref{eq:SlimSigmar0} provides the prescription for the correct matched filter (only)
in the limit of $\sigma_{r}\rightarrow0$.


\subsection{The PSF of the difference image}

\begin{figure}
\centerline{\includegraphics[width=8.5cm]{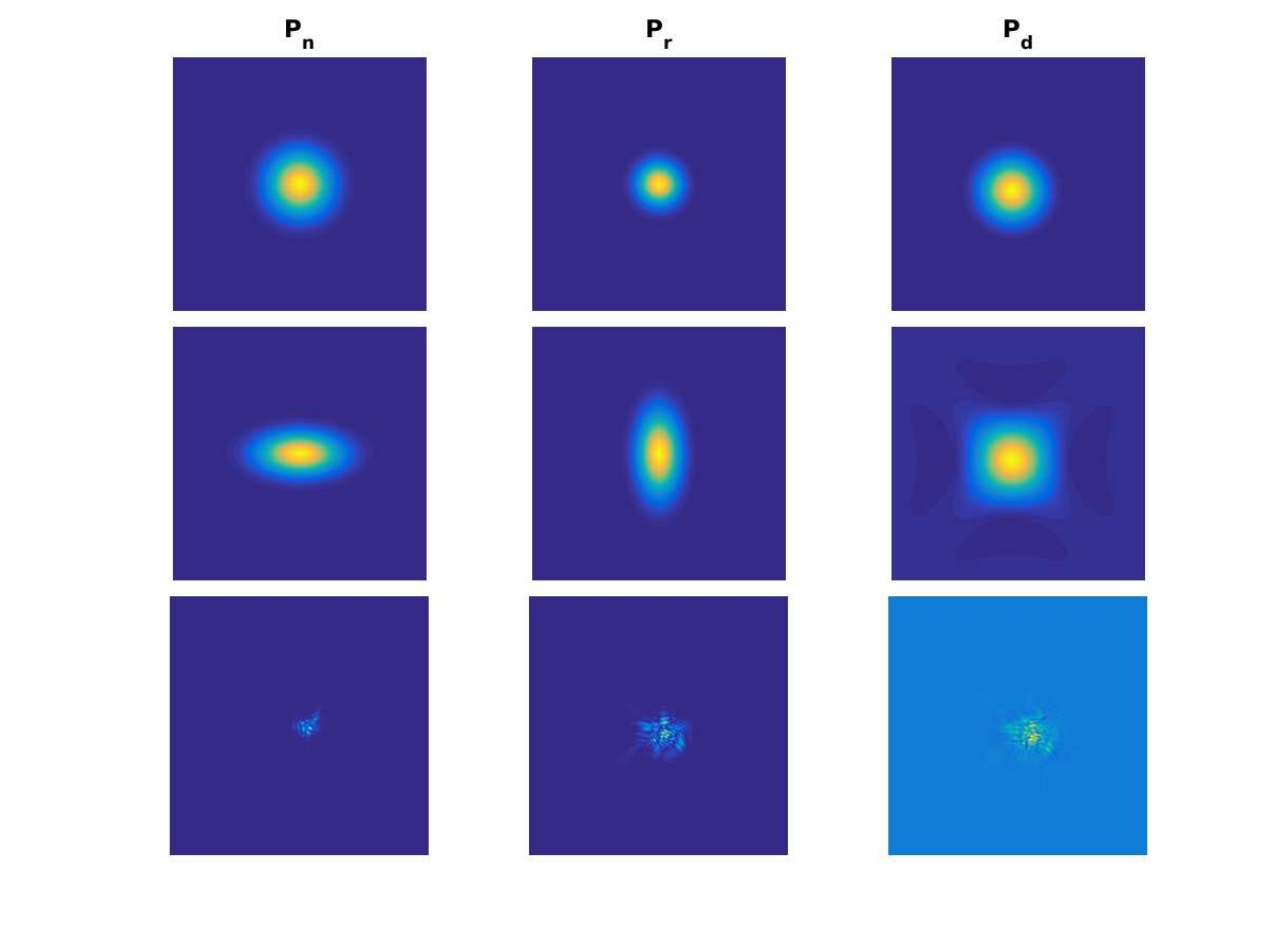}}
\caption{$P_{n}$ (left column), $P_{r}$ (middle column)
and the corresponding
$P_{D}$ (right column) for three cases.
The first row is for the case of symmetric Gaussian PSFs
with sigma-width of 2 and 3\,pix for the new and reference, respectively.
The second row is for the case of a-symmetric Gaussian PSFs
with sigma-width of 2 by 4\,pix and 4 by 2\,pix for the new and reference, respectively.
In the third row $P_{n}$ and $P_{r}$ are simulated
speckle images (using the tools in Ofek 2014).
In the speckle simulations we set $D_{tel}/r_{0}$=20,
where $D_{tel}$ is the telescope diameter and $r_{0}$ is the Fried length.
\label{fig:PnPrPd}}
\end{figure}

The PSF, $P_{D}$, of the proper subtraction image is a combination
of $P_{n}$ and $P_{r}$.
In Figure~\ref{fig:PnPrPd} we present $P_{n}$, $P_{r}$ and the corresponding
$P_{D}$ for three cases, of symmetric Gaussians,
a-symmetric Gaussians, and speckle images.

\subsection{Knowledge of the PSFs}

An apparent drawback of our method is that one needs to know
the PSFs of the images, while in the \citet{AL98}
family of methods one simply solves
for the convolution kernel $\widehat{P_{n}}/ \widehat{P_{r}}$
without measuring the PSFs.

However, one can write the expression for $D$ with
$\widehat{P_{n}}/\widehat{P_{r}}$,
allowing to incorporate relative knowledge of the PSFs.
This is relevant in rare cases of images that
contain no point sources, for example, only galaxies.
%
However, in order to optimally find transients in the image
all the methods requires the PSFs (see \S\ref{sec:NoislessR}).
In any case, in most observational situations the PSF is measurable from point sources in the
image and therefore this should not be considered as a drawback.

%


\subsection{Registration and color-refraction errors}
\label{eq:RegErr}

Image subtraction relies on many steps taken prior to the
differencing process.
Any noise introduced by the pre-processing steps
will be propagated into the final subtraction image.
Examples for such problems include:
registration errors,
color-refraction systematic errors,
and small-scale flat-fielding errors.

Here we suggest two types of treatments for such noise:
(1) It is straightforward to introduce these extra sources
of noise into the variance image of $S$ and use it to
calculate $S_{{\rm corr}}$
(see \S\ref{sec:sim} for examples).
This correction is sub-optimal, but it is resilient
to pre-processing errors.
(2) An accurate treatment of the problem is to fit any astrometric shift
and flux variation for each detected artifact in the difference image $D$ (see \S\ref{sec:AstNoise}).
Albeit this is computationally expensive, this kind of solution
is very common in astronomy
(e.g., {\tt DAOPHOT}, \citealp{Stetson1987}; {\tt DOPHOT}, \citealp{Schechter1993}).

In the future, it is possible that the use of these steps may enable us to remove completely the need for any post-subtraction transient identification
using machine learning or human classification.
We note that successful implementation of these ideas requires good understanding of all the sources of noise.

\subsection{Free parameters}

In principle, our method does~not have any free parameters that the user
needs to set.
We note that the \cite{AL98}, \cite{Bramich}, and \cite{Yuan08} methods do have internal
degrees of freedom
that the user needs to define and that may influence the final outcome.
For example, the \cite{Bramich} method may be sensitive to the kernel size,
while the \cite{AL98} method depends on the basis functions one chooses to represent
the convolution kernel (see e.g., \citealp{Becker2012} and \citealp{Bramich2015}).

\subsection{Computational complexity}

In terms of computational complexity, our subtraction method is fast, as the most demanding operation in our image subtraction method is the FFT operation
(or alternatively convolution with a small kernel).
Tests indicate that our algorithm is at least an order
of magnitude faster than the inversion algorithms
by \cite{AL98} and \cite{Bramich} as they are essentially solving
a linear least square problem with a large number of equations
and tens to hundreds of unknowns.

\section{Summary of algorithm}
\label{sec:algo}
We recommend to perform the subtraction on small image patches
in order to minimize residual astrometric shifts, inhomogeneous transparency and background.
In addition, it allows to use position-dependent PSFs.
The image patches should be overlapping by at least two PSF lengths, in each dimension,
in order to avoid edge effects of the convolution process.

A step-by-step outline of our algorithm is as follows:\\
\\
{\bf Input arguments:}\\
$N$ - background subtracted new image (registered to $R$).\\
$R$ - background subtracted reference image.\\
$N_{b}$ - new image including background in electron units.\\
$R_{b}$ - reference image including background in electron units.\\
$P_{n}$ - PSF of new image normalized to have unit sum.\\
$P_{r}$ - PSF of reference image normalized to have unit sum.\\
$\sigma_{n}$ - std of the background of the new image.\\
$\sigma_{r}$ - std of the background of the reference image.\\
$r_{n}$ - read noise of new image in electrons.\\
$r_{r}$ - read noise of reference image in electrons.\\
$\sigma_{x}$ - rms (in pixels) of the astrometric registration solution in the X-axis.
This is either a scalar or a matrix.\\
$\sigma_{y}$ - rms (in pixels) of the astrometric registration solution in the Y-axis.
This is either a scalar or a matrix.\\
\\
{\bf Output:}\\
$D$ - The proper difference image.\\
$P_{D}$ - The PSF of the proper difference image.\\
$S_{{\rm corr}}$ - The matched filter difference image corrected for source noise
and astrometric noise.\\
$P_{D_n}$ - The PSF of a delta function in $N$ as it appears in $D$.\\
$P_{D_r}$ - The PSF of a delta function in $R$ as it appears in $D$.\\
\\
{\bf Algorithm:}\\
\begin{enumerate}
  \item Optionally construct a reference image ($R$; Eq.~\ref{eq:ProperCoadd}),
        its PSF ($P_{r}$; Eq.~\ref{eq:ProperCoaddPSF})
        and flux ($F_{r}$; Eq.~\ref{eq:Fr})
        using the Zackay \& Ofek (2015b)
        proper coaddition method.
  \item Solve Equation~\ref{eq:DnDrBeta} for the best-fit value of
        $\beta$ and optionally $\gamma$, $\Delta{x}$, and $\Delta{y}$
        (need to use Eqs.~\ref{eq:Dn} and \ref{eq:Dr}).
        Since this Equation is non-linear in $\beta$ use iterations.
        Set $\beta=1$ in the first iteration, update
        the value of $\beta$ and continue until convergence.
        Use robust fitting\footnote{Robust fitting is less sensitive
to outliers. An example for a robust fitter is the {\tt robustfit.m} function in MATLAB.}.
  \item If applicable calculate $\gamma$ (Equation~\ref{eq:gammatag}) and subtract $\gamma$ from $N$.
  \item If applicable, shift $P_{n}$ by $\Delta{x}$ and $\Delta{y}$.
  \item Set $F_{r}=1$ and $F_{n}=\beta$.
  \item Calculate $\widehat{D}$ (Eq.~\ref{eq:D}).
  \item Calculate $\widehat{P_{D}}$ (Eq.~\ref{eq:Dpsf}).
  \item Calculate $\widehat{S}=\overline{\widehat{P_{D}}} \widehat{D}$.
  \item Calculate $\widehat{P_{D_n}}$ (Eqs.~\ref{eq:Pdn} and \ref{eq:FdN}).
  \item Calculate $\widehat{P_{D_r}}$ (Eqs.~\ref{eq:Pdr} and \ref{eq:FdR}).
  \item Calculate $k_{r}$ (Eq.~\ref{eq:kr}).
  \item Calculate $k_{n}$ (Eq.~\ref{eq:kn}).
  \item Set $V(\epsilon_{n}) = N_{b}+r_{n}^{2}$ and calculate $V(S_{N})$ (Eq.~\ref{eq:VSN}).
  \item Set $V(\epsilon_{r}) = R_{b}+r_{r}^{2}$ and calculate $V(S_{R})$ (Eq.~\ref{eq:VSR}).
        If $R$ is composed of multiple images, it is better
        to sum up the $V(S_{R_{j}})$ of the individual
        reference images (see Appendix~\ref{Ap:CorrectSourceNoise} and Eq.~\ref{eqA:VSRj}).
  \item Calculate $V_{{\rm ast}}(S_{N})$ (Eqs.~\ref{eq:VastSN} and \ref{eq:SN}).
  \item Calculate $V_{{\rm ast}}(S_{R})$ (Eqs.~\ref{eq:VastSR} and \ref{eq:SR}).
  \item Calculate $S_{{\rm corr}}$ (Eq.~\ref{eq:Scorr}).
        As a sanity check, the (robust) std of $S_{{\rm corr}}$ should be $\approx 1$.
  \item Search for local maxima in $S_{{\rm corr}}$ - the peak value corresponds to the significance of the transient in units of sigmas.
  \item As an alternative to steps 15 and 16, we can search all locations in $D$
        that correspond to statistically significant sources
        in $S_{{\rm corr}}$ (without astrometric contributions) for moving point sources using $P_{S}$ (Eq.~\ref{eq:shift}),
        measure their flux and astrometric variability and subtract them.
  \item Select remaining sources with significance larger than some threshold, determined from the desired false alarm probability.
  \item Calculate the flux of the transient candidates using Equations~\ref{eq:FluxEst}, \ref{eq:F_S}, and \ref{eq:FluxStdEst}.
\end{enumerate}


\section{Tests}
\label{sec:tests}

There are several important challenges in testing any image differencing algorithm.
Image subtraction in general is affected by many factors.
%
%
Therefore, it is desirable to separate between
external problems (e.g., non-perfect registration)
and issues related to the subtraction itself (e.g., numerical stability).
Therefore, we are using both simulations and real data to test our
image differencing algorithm.

It is worthwhile to compare the new algorithm with existing methods.
However, such a comparison is problematic, as other methods
do~not specify the matched filter for source detection.
Furthermore, some of these methods depend on the selection of basis
functions and kernel size.
In addition, there are several ways to solve a system
of linear equations (e.g., SVD) and these may
influence the final outcome.
Therefore, here our comparison with other methods is limited.

In \S\ref{sec:sim} we present tests based on simulated data,
while in \S\ref{sec:real} we discuss real images.
The code we use is available as part of the
Astronomy and Astrophysics package for MATLAB (\citealp{Ofek2014}), described in \S\ref{sec:code}.

\subsection{Simulations}
\label{sec:sim}

An important feature of our algorithm is its numerical stability.
The best way to test this is on simulations, as the input is fully controlled.

\begin{figure*}
\centerline{\includegraphics[width=16cm]{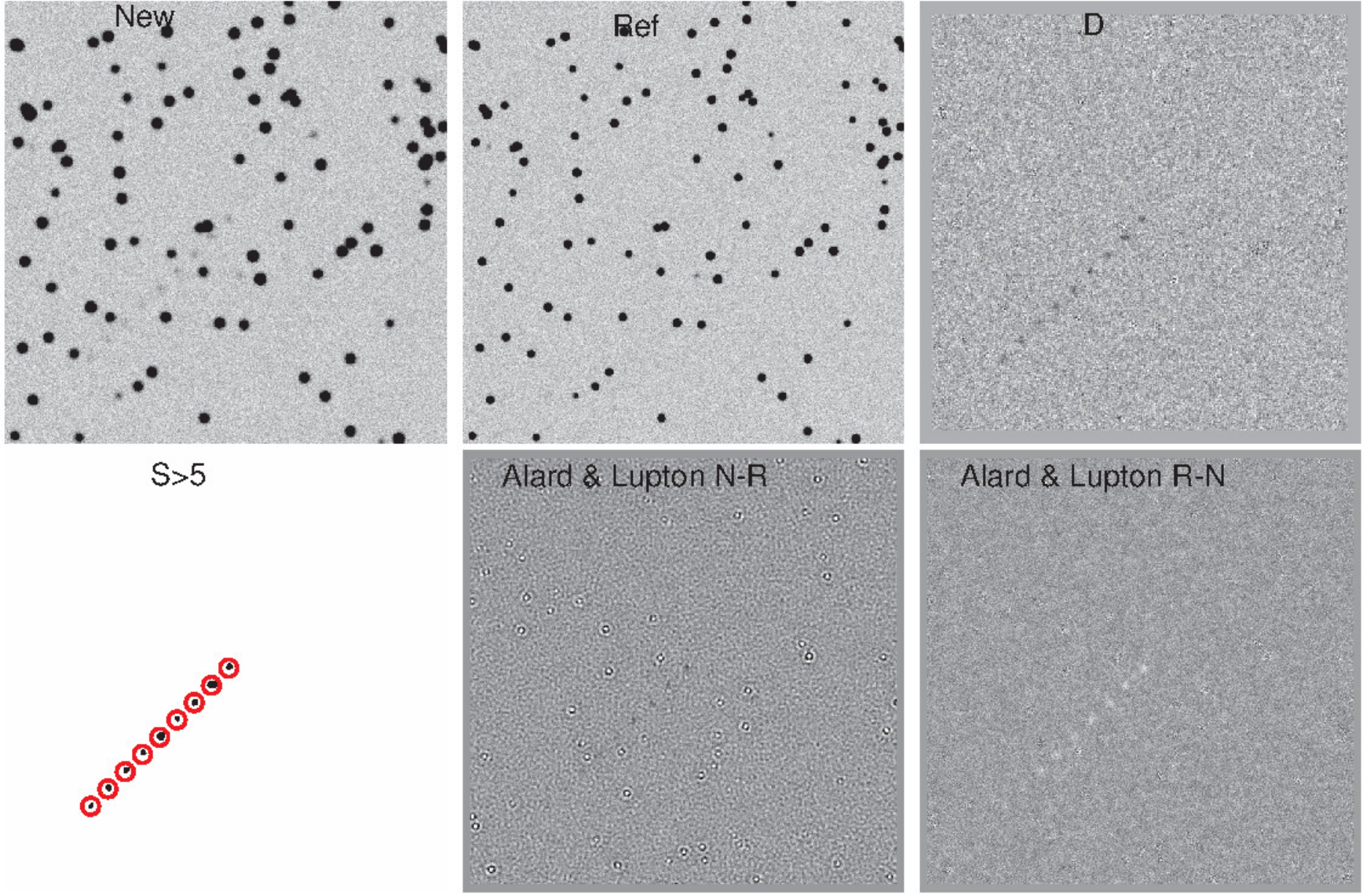}}
\caption{Subtraction of simulated images with
symmetric Gaussian PSF with
sigma-width of 2 and 3\,pixels, for the reference and new images, respectively.
Left to right (top):
the new image, the reference image, the proper subtraction image ($D$);
(bottom)
the matched-filtered image ($S$) with 5-$\sigma$ threshold,
the \cite{AL98} subtraction of new minus reference,
and the \cite{AL98} subtraction of reference minus new.
The position of the simulated transient sources in the thresholded
matched filtered image are marked by red circles.
All the images are presented with inverted grayscale map.
\label{fig:Sim2}}
\end{figure*}

We simulated images of $512$ by $512$ pixels size, with background level of 300\,electrons,
with Poisson noise.
In each image we simulated 100 stars with integrated flux taken from a flat distribution
between 0 to $10^{5}$\,electrons and Poisson noise.
In addition we added to the new image nine transient sources with position and flux as listed in Table~\ref{tab:PosFlux}.
In the first set of simulated images the PSF of the sources in the images are symmetric Gaussians with
sigma-width of 2 and 3\,pixels, for the reference and new images, respectively.
Figure~\ref{fig:Sim2} shows, left to right (top):
new image, the reference image, the proper subtraction image ($D$);
(bottom) the matched-filtered image ($S$) threshold above 5-$\sigma$,
the \cite{AL98} subtraction of new minus reference,
and the \cite{AL98} subtraction of reference minus new.
The \cite{AL98} subtractions are based on the ISIS software (Alard \& Lupton 1999).
This Figure demonstrates that while our image subtraction method is symmetric,
the \cite{AL98} algorithm is not symmetric.
In this case it is working well in one direction, but subtraction artifacts
are clearly visible (ringing due to deconvolution) in the other direction.
Furthermore, thresholding our matched filter image above 5-$\sigma$ 
reveals only the simulated transients.
\begin{deluxetable}{lll}
\tablecolumns{3}
\tablewidth{0pt}
\tablecaption{Simulated transients in the new image}
\tablehead{
\colhead{X}               &
\colhead{Y}        &
\colhead{Flux}    \\
\colhead{(pix)}           &
\colhead{(pix)}            &
\colhead{(electrons)}
}
\startdata
100 &  100  &  1500 \\
120 &  120  &  1600 \\
140 &  140  &  1800 \\
160 &  160  &  2000 \\
180 &  180  &  2200 \\
200 &  200  &  2400 \\
220 &  220  &  2600 \\
240 &  240  &  2800 \\
260 &  260  &  3000 
\enddata
\tablecomments{The position and mean flux of simulated transient sources in the
new images in Figures~\ref{fig:Sim2}--\ref{fig:Sim3}.}
\label{tab:PosFlux}
\end{deluxetable}

\begin{figure*}
\centerline{\includegraphics[width=16cm]{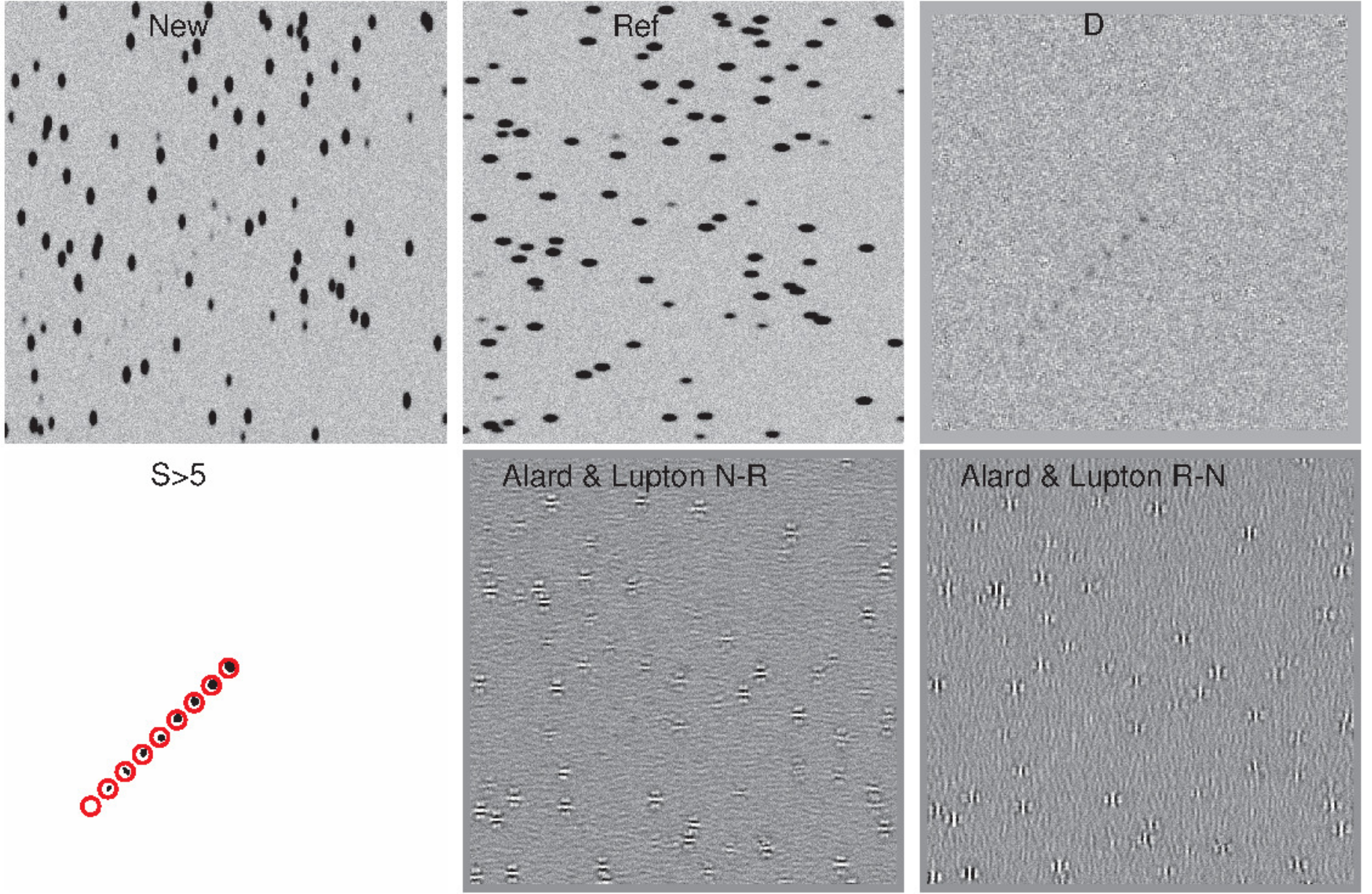}}
\caption{The same as Figure~\ref{fig:Sim2}, but for the subtraction of simulated images with
a-symmetric Gaussian PSF with
sigma-width of 2 by 4\,pix in the new image
and 4 by 2\,pix in the reference image.
\label{fig:Sim1}}
\end{figure*}

Next we simulated images with the same parameters as in
Figure~\ref{fig:Sim2},
with a-symmetric Gaussian PSF with sigma 
width of 2 by 4\,pix in the new image
and 4 by 2\,pix in the reference image.
Figure~\ref{fig:Sim1}, is the same as Figure~\ref{fig:Sim2}, but for these images.
Again, the a-symmetry of the \cite{AL98} family of methods is
seen. Furthermore, in this case the ringing due to deconvolution
is seen in both the $N-R$ and $R-N$ subtractions.

\begin{figure*}
\centerline{\includegraphics[width=16cm]{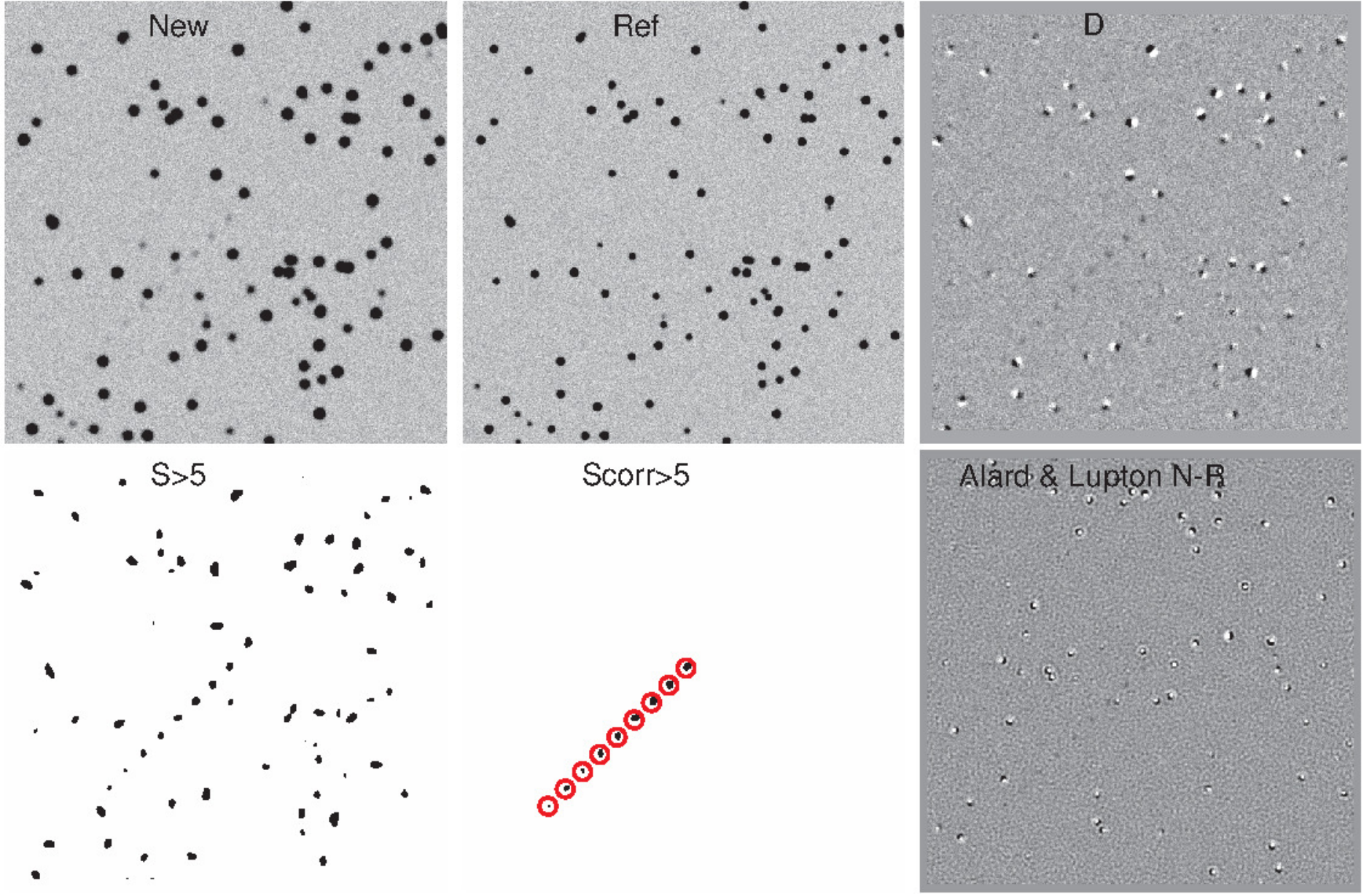}}
\caption{Subtraction of simulated images with 0.3\,pix (rms) astrometric noise and
symmetric Gaussian PSF with
sigma-width of 2 and 3\,pixels, for the reference and new images, respectively.
Left to right (top):
the new image, the reference image, the proper subtraction image ($D$).
Left to right (bottom):
the matched-filtered image ($S$) threshold above 5-$\sigma$,
the source noise corrected and astrometric noise corrected
matched-filtered image ($S_{{\rm corr}}$) threshold above 5-$\sigma$,
and the \cite{AL98} subtraction of the new minus reference.
The position of the simulated transient sources in the thresholded
matched filtered image are marked by red circles.
\label{fig:Sim3}}
\end{figure*}

One of the most important practical features of our new method is the ability
to incorporate other types of noise into the detection process
(e.g., source noise, astrometric noise, color-refraction noise).
To demonstrate this we repeated the first simulation (Figure~\ref{fig:Sim2}),
but this time with normally distributed astrometric noise with standard deviation
of 0.3\,pix.
Figure~\ref{fig:Sim3} shows, left to right (top):
the new image, the reference image, the proper subtraction image ($D$);
(bottom):
the matched-filtered image ($S$) thresholded above 5-$\sigma$,
the source noise corrected and astrometric noise corrected
matched-filtered image ($S_{{\rm corr}}$) thresholded above 5-$\sigma$,
and the \cite{AL98} subtraction of the new minus reference.
In this case, the subtraction contains a large number
of positive-negative residuals, but our $S_{{\rm corr}}$
image deals well with this astrometric noise,
and only the simulated transients are detected.

\subsection{Tests on real images}
\label{sec:real}

We tested the new method on imaging data available
from the Palomar Transient Factory (PTF\footnote{http://www.ptf.caltech.edu/iptf}; Law et al. 2009; Rau et al. 2009)
data release 2.
The image processing is described in \citet{Laher2014}
while the photometric calibration is discussed in \citet{Ofek2012}.

Table~\ref{tab:PosFlux} lists the various images on which we tested our algorithm.
Registration, background subtraction and PSF estimation
were performed using the code described in \S\ref{sec:code}.
\begin{deluxetable*}{llllllll}
\tablecolumns{7}
\tablewidth{0pt}
\tablecaption{List of tests on real images}
\tablehead{
\colhead{Test}               &
\colhead{Field/CCD}               &
\colhead{Size}        &
\colhead{$N$}    &
\colhead{$R$}    &
\colhead{$FWHM_{N}$}    &
\colhead{$FWHM_{R}$}    \\
\colhead{}               &
\colhead{}               &
\colhead{(pix)}        &
\colhead{}    &
\colhead{}    &
\colhead{(arcsec)}    &
\colhead{(arcsec)}    
}
\startdata
1     & 100031/04  & $560\times560$   & 2012-12-20.4134   & proper          &   5.4   &  2.9 \\  
2     & 100031/11  & $1000\times1000$ & 2011-08-08.1839   & 2011-04-12.1865 &   2.5   & 2.9    
\enddata
\tablecomments{List of tests on real images. ``proper'' indicates a reference image that was constructed using proper coaddition (Zackay \& Ofek 2015b).}
\label{tab:RealTests}
\end{deluxetable*}

\begin{figure*}
\centering
\includegraphics[width=16cm]{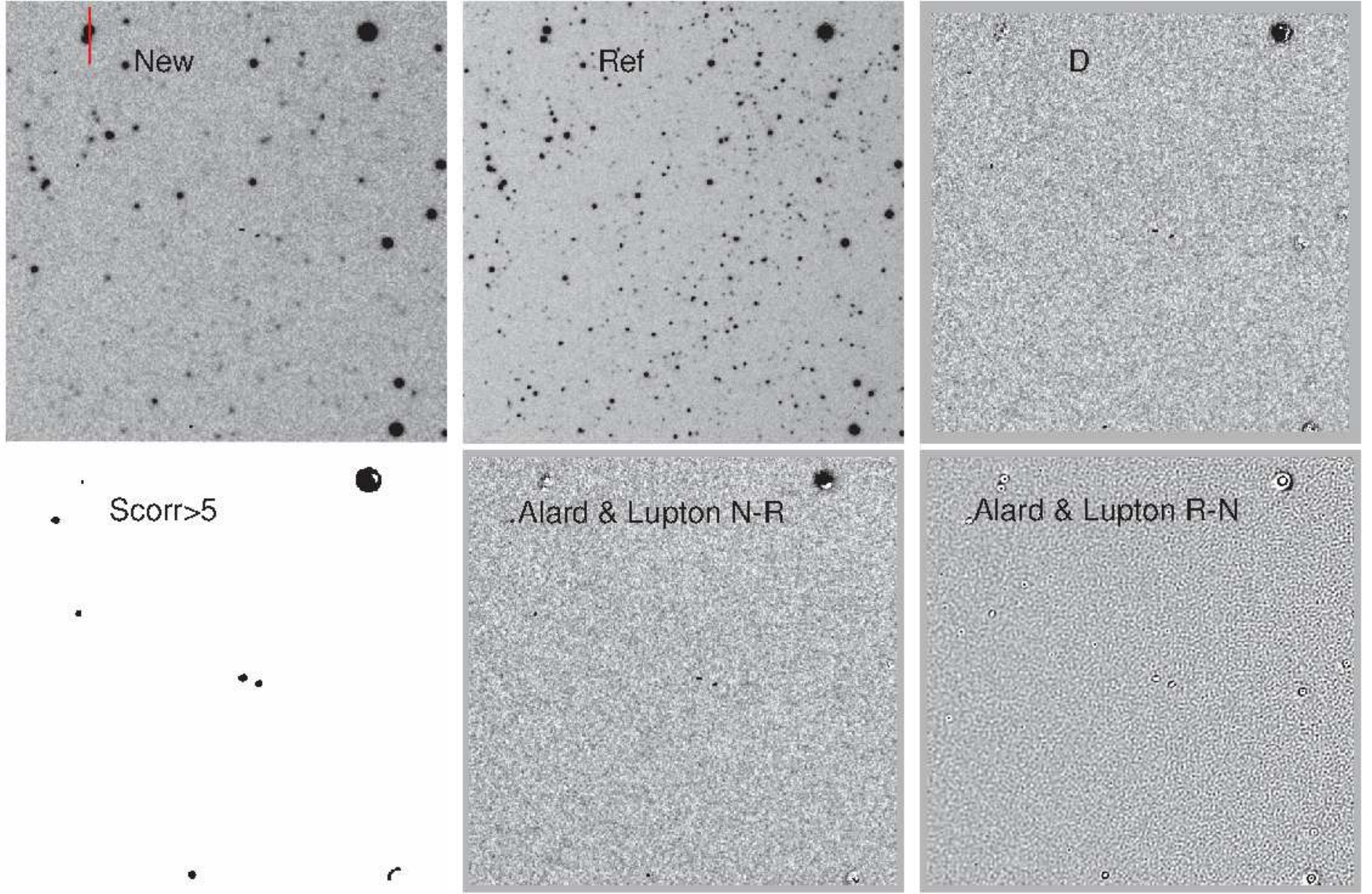}
\caption{Image subtraction results for test 1 (Table~\ref{tab:RealTests}). Left to right (top): the new image, the reference image,
the proper subtraction image $D$; (bottom) the matched filter corrected difference image ($S_{{\rm corr}}$) filtered at 5-$\sigma$, the \cite{AL98} ISIS subtraction of the new minus the reference, and the ISIS subtraction of the reference minus the new. All the images are presented with inverted grayscale map. The Red line (in the new panel) indicates the position of the profile cut we present in Figure~\ref{fig:Test1a_cut}. In the $S_{{\rm corr}}>5$ map, CR1--CR5 indicate the position of cosmic rays detected by our algorithm, while the two bright residuals on the right part of the image are due to saturated stars.
The residual at the top left has a significance of 5.7-$\sigma$ and it is at the interface between two bright stars. The mechanism that generate this particular residual is discussed in \S\ref{sec:Add}. \label{fig:Test1a}}
\end{figure*}

Figure~\ref{fig:Test1a} presents the image subtraction results of test 1.
The top panels left to right are: the new image, the reference image
and the proper difference image $D$.
The bottom panels left to right are:
the matched filter corrected image ($S_{{\rm corr}}$) thresholded at 5-$\sigma$,
the \cite{AL98} subtraction of the $N-R$,
and the \cite{AL98} subtraction of the $R-N$.
Figure~\ref{fig:Test1a} also demonstrates
that the \cite{AL98} subtraction is not symmetric to the exchange of $R$ and $N$, while our method is.
Specifically, the $R-N$ image of the \cite{AL98}
has strong, and high amplitude, correlated noise.

\begin{figure}
\centerline{\includegraphics[width=8cm]{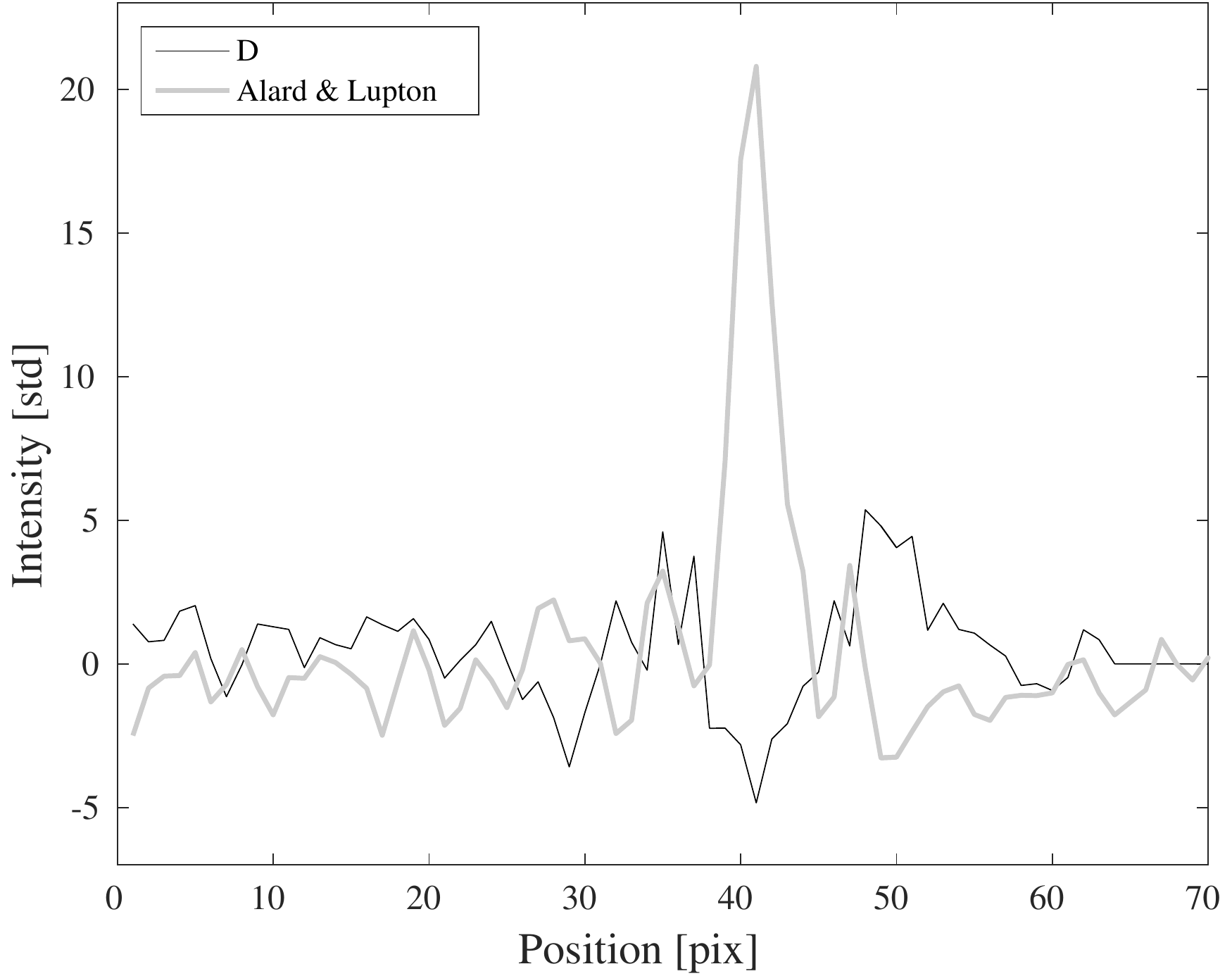}}
\caption{A profile cut, at the position
of the red line in Figure~\ref{fig:Test1a},
in the proper subtraction image $D$ (black line)
and the \cite{AL98} subtraction ($N-R$; gray line).
The images are normalized such that the standard deviation
of the images is unity.
This demonstrates that in the presence of bright stars,
the fluctuations in our subtraction image
are modest, while the residuals
in the \cite{AL98} subtractions are large.
We note that the $D$ image is not filtered while
the \cite{AL98} subtraction is partially filtered.
Therefore, the noise properties of $D$, relative
to the \cite{AL98} subtraction, are even
better than indicated from this plot.
\label{fig:Test1a_cut}}
\end{figure}

On first glance, the \cite{AL98} $N-R$ image looks cosmetically good.
However, on closer inspection we can see that this image
has subtraction residuals with large amplitude.
For example, Figure~\ref{fig:Test1a_cut} shows a profile cut, at the location
of the red line in Figure~\ref{fig:Test1a},
in the proper subtraction image $D$
and the \cite{AL98} subtraction ($N-R$).
The images are normalized such that the standard deviation
of the images is one.
This Figure clearly shows that while our algorithm
behaves very well in the presence of stars,
the \cite{AL98} subtraction has very large fluctuations.
We note that the fact that the \cite{AL98} subtraction image
is partially filtered is seen by eye (i.e., smoother noise)

\begin{figure*}
\centerline{\includegraphics[width=16cm]{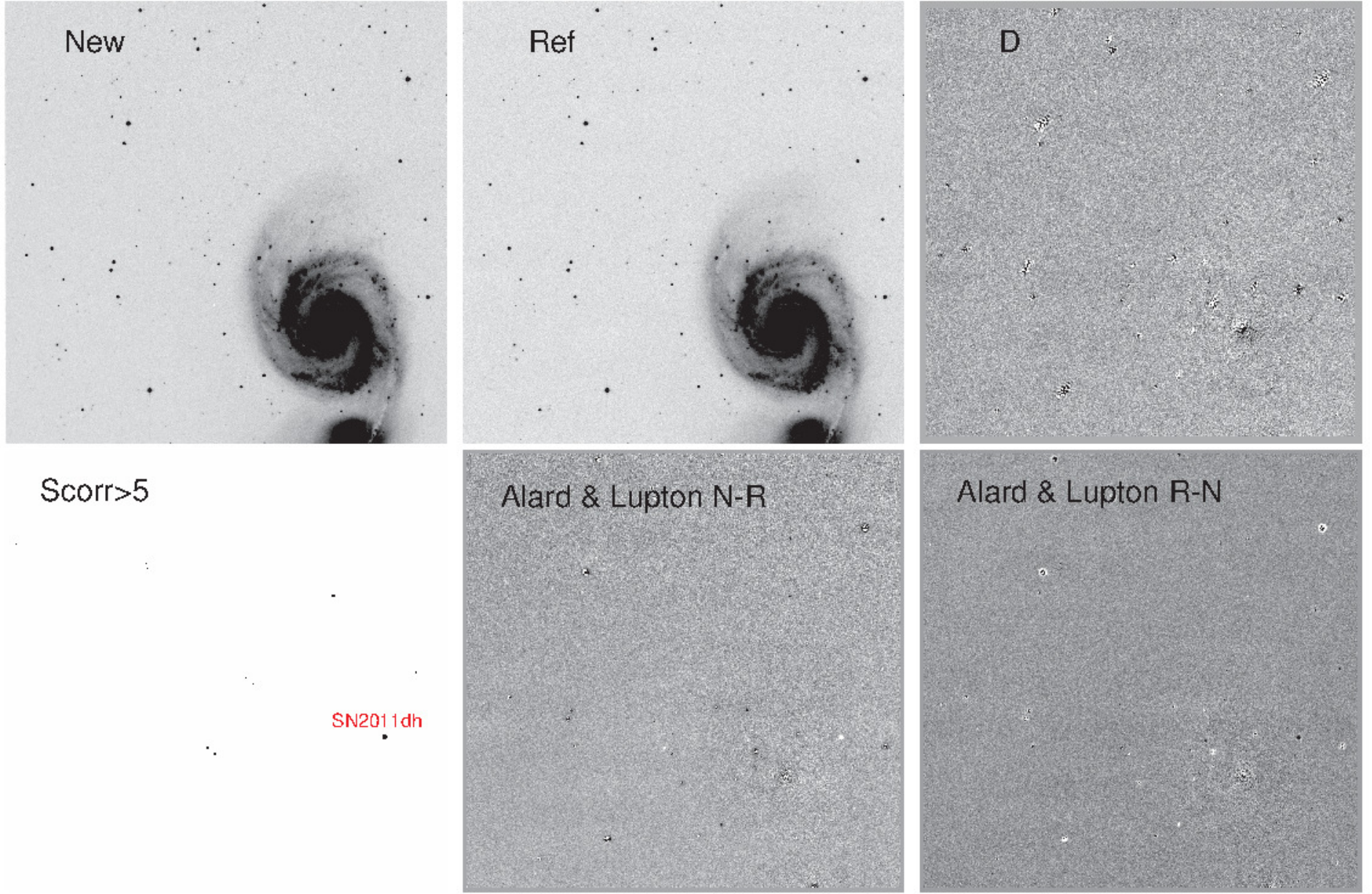}}
\caption{The same as Figure~\ref{fig:Test1a}, but for
the test 2 images, containing the bright
galaxy M51 and SN\,2011dh.
The detected sources in $S_{{\rm corr}}>5$ are SN\,2011dh,
particle hits and bad pixels.
\label{fig:Test3a}}
\end{figure*}

Figure~\ref{fig:Test3a} is the same as Fig.~\ref{fig:Test1a},
but for the subtraction of images of test 2.
These images contain the bright galaxy M51,
and SN\,2011dh (\citealp{Arcavi2011}).
We note that in $D$ we clearly see residuals due to mis-alignment
of the images. However, these residuals are gone
when we present $S_{{\rm corr}}$ that takes the astrometric noise
into account.
We further note that the astrometric residuals are less pronounced
in the \cite{AL98} subtraction simply because these images
are partially filtered, and therefore smoother.
The transient candidate detected in the
$S_{{\rm corr}}$ image above 5-$\sigma$ threshold are SN\,2011dh,
cosmic rays and bad pixels.

\section{Code}
\label{sec:code}

We present two sets of codes based on MATLAB and Python.
The MATLAB code contains functions to deal
with all the image processing steps, including the
registration and PSF estimation.
The MATLAB code is available as part of the
MATLAB Astronomy and Astrophysics package\footnote{http://webhome.weizmann.ac.il/home/eofek/matlab/} (Ofek 2014).
This code is under development and we expect that improved versions will be available in the future.
The Python code\footnote{https://sites.google.com/site/barakzackayhomepage/} contains only a simple implementation of our algorithm
that requires as input: fully registered images, as well as their PSF,
background images and variance images.

The main high-level MATLAB functions required for image subtraction are listed in Table~\ref{tab:Fun}
along with their brief description.
Some of these functions are discussed in Zackay \& Ofek (2015a, 2015b).
The implementation details related to some of these utilities are further discussed in \S\ref{sec:Details}.
\begin{deluxetable*}{lll}
\tablecolumns{2}
\tablewidth{0pt}
\tablecaption{High-level functions relevant for coaddition}
\tablehead{
\colhead{Name}               &
\colhead{Description  }    \\
\colhead{}           &
\colhead{}
}
\startdata
{\tt imsub\_fft.m}             & Optimal subtraction of two images.\\
                               & The function optionally registers the images, finds the PSF, background, \\
                               & and variances. The function returns $D$, $S$, $S_{{\rm corr}}$, $P_{D}$, \\
                               & $P_{D_{r}}$ and $P_{D_{n}}$. Upon user request, $S_{{\rm corr}}$ may be corrected \\
                               & for astrometric noise and color-refraction noise. \\
\hline
{\tt psf\_builder.m}           & Construct a PSF template by re-sampling the pixels around \\
                               & selected bright/isolated stars.\\
{\tt sim\_coadd.m}             & Coadd a list of images, using various weighting schemes.\\
                               & The function also allows for filtering the images prior to the coaddition.\\
                               & The function can also align the images, calculate the weights and PSFs.\\
{\tt sim\_coadd\_proper.m}     & Proper coaddition of images (see Zackay \& Ofek 2015b). \\
                               & The function can also align the images, calculate the weights and PSFs.\\
{\tt sim\_align\_shift.m}      & Register a set of images against a reference image.\\
                               & The function assumes the images can be registered \\
                               & using an arbitrary large shift, but only a small rotation term.\\
{\tt weights4coadd.m}          & Calculate parameters required for calculation of weights for \\
                               & coaddition. Including the background, its variance, estimate of the \\
                               & flux-based zero points (i.e., transparency), and measure the PSF.\\
{\tt sim\_back\_std.m}         & Estimate the spatially-dependent background and variance of images.
\enddata
\tablecomments{High-level functions relevant for coaddition, which are part of the
Astronomy and Astrophysics toolbox for MATLAB (Ofek 2014).}
\label{tab:Fun}
\end{deluxetable*}

\section{Implementation details}
\label{sec:Details}

Given background subtracted images, their variance, PSF
and flux-based zero points ratio,
our image subtraction method is presented using closed-form formula.
Therefore, the implementation of this method is simple and rigorous,
and does~not require special attention.
However, like any other method for image subtraction,
this technique is sensitive to the steps taken prior
to the image subtraction (e.g., registration).

Here we discuss some of the details that can greatly influence
the successful application of any image subtraction algorithm.

\subsection{Background and variance estimation}

The background and variance in real wide-field-of-view
astronomical images cannot be treated as constants over the entire field of view.
Therefore, we suggest to estimate them locally and interpolate.
To estimate the background and variance one needs to make sure that the
estimators are not biased by stars or galaxies.
Following Zackay \& Ofek (2015a, 2015b) we suggest to fit a Gaussian to the histogram of the image pixels
in small regions\footnote{We are currently using $256\times256$~arcsec$^{2}$ blocks.},
and to reject from the fitting process pixels with high
values (e.g., the upper 10\% of pixel values).
Regions containing large galaxies or complex background
may require special treatment.

\subsection{PSF estimation and spatial variations}
\label{sec:PSF}

We note that Equations~\ref{eq:S} and \ref{eq:D} are roughly linear
to perturbations in the PSF, compared with the real PSF.
Among the complications that may affect the PSF measurement
are pixelization, interpolation and the resampling grid.
Furthermore, the PSF is likely not constant spatially and it also
may change with intensity due to charge self repulsion.
This specifically may lead to the brighter-fatter effect (e.g., Walter 2015).

In some cases the PSF may vary over the field of view.
The simplest approach is to divide the image to smaller images
in which the PSF is approximately constant.
These sub-images can be as small as four times the PSF size.
Since the convolution operation is local, it is straight forward
to incorporate a spatially variable PSF into any
subtraction method (e.g., \citealp{Alard2000}).

\subsection{Interpolation}

The registration step requires to interpolate one of the images
into a new coordinates grid.
If the PSF is Nyquist sampled (band limited) then one can use
the Whittaker-Shannon interpolation formula (sometimes called sinc interpolation)
without losing information due to the interpolation process.

However, if the PSF is undersampled, interpolation will lead
to variation in the PSF shape which depends on the position
of the source within the pixel (pixel phase).
Such an effect may cause severe problems to any subtraction
method.
One simple way to deal with this problem is to add a noise
term to the denominator of $S_{{\rm corr}}$ (Eq.~\ref{eq:Scorr})
that takes into account the extra noise induced by the pixel-phase dependent
PSF variations.
Such a correction is under development.

\subsection{Registration}

Registration is a critical step for any image differencing technique.
Any leftover registration imperfection residuals between
the new and reference image will lead to improper subtraction,
subtraction artifacts and eventually to false detections.
In \S\ref{sec:SourceNoise} and \ref{sec:AstNoise}
we discuss how registration errors,
color-refraction and astrometric scintillations can be treated.
However, it is still desirable to minimize any registration
errors prior to subtraction.

In many cases affine transformations are not enough to map
between the two images.
The main reasons include:
differential atmospheric refraction,
differential aberration of light, and
high-order optical distortions.

Usually when images are taken with the same system
and the same on-sky pointing, optical distortions will not play an important role
as their effect on the two images is almost identical.

\begin{figure}
\centerline{\includegraphics[width=8.5cm]{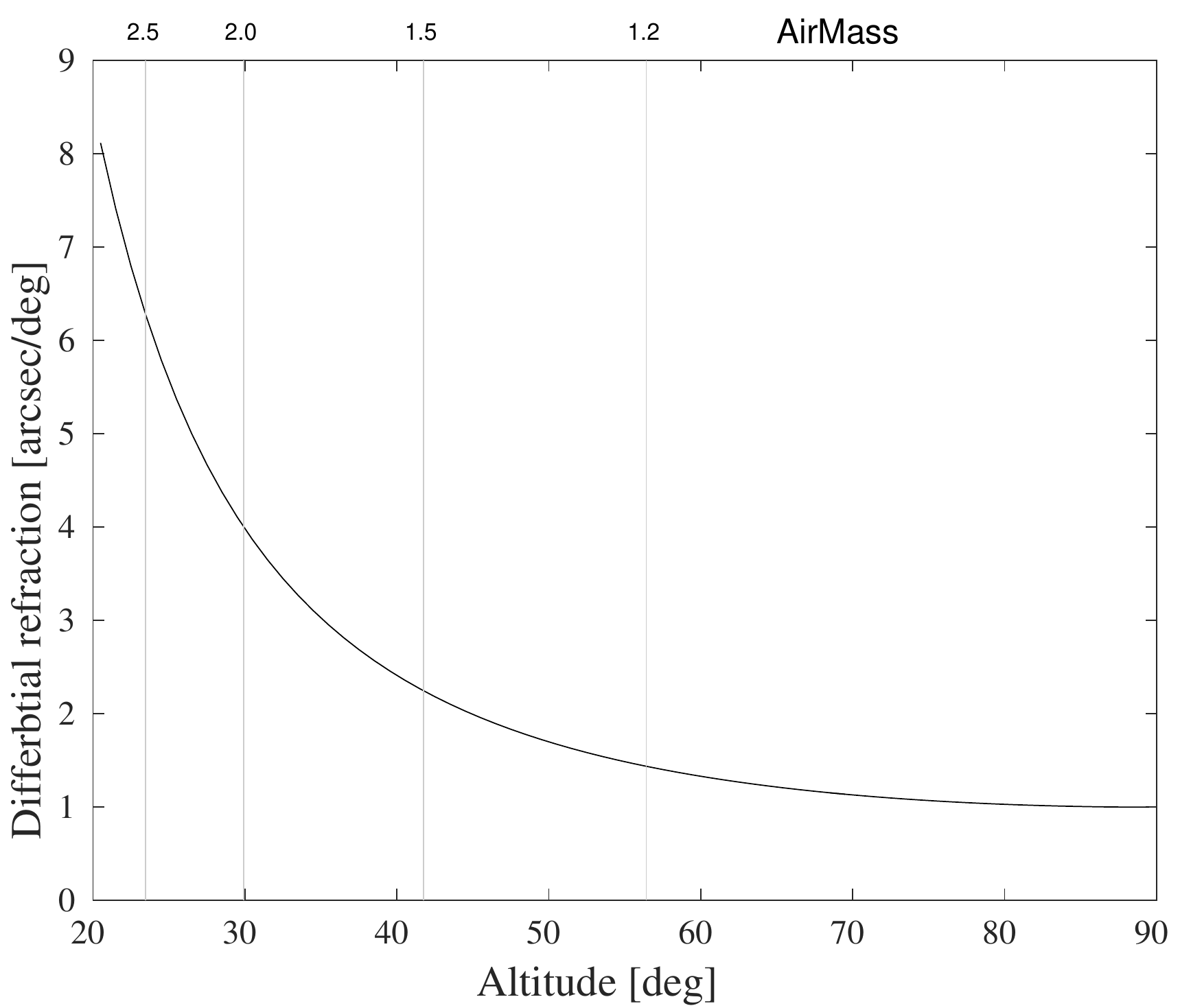}}
\caption{The differential atmospheric refraction (in the altitude direction),
in units of arcsec per deg.
Calculated using the code in \cite{Ofek2014} and formulae provided in Filippenko et al. (1982),
for a wavelength of 5000\,\AA, temperature of $15^{\circ}$\,C,
pressure of 760\,mm\,Hg and partial water vapor
pressure of 8\,mm\,Hg.
\label{fig:DiffRef_Alt}}
\end{figure}

The amplitude of differential atmospheric refraction
can be as high as $8''$\,deg$^{-1}$. Figure~\ref{fig:DiffRef_Alt} shows
the amplitude of differential atmospheric refraction
as a function of altitude.
Since the direction of the atmospheric refraction
is known very well, the best way
to deal with the distortions caused by the atmosphere
is to add to the affine transformation
terms that fit the atmospheric refraction amplitude with its known direction
(i.e., the parallactic angle).
%
Unfortunately, most astrometric and registration packages
do~not support distortions of this form,
and instead they absorb the refraction correction into high
order polynomials.
Furthermore, the current WCS header keywords do~not support
this kind of transformations.
Our code described in \S\ref{sec:code} does support this transformation.

We note that atmospheric refraction distortions are detectable
even on small angular scales. For example, this effect can reach 
$0.1''$\,arcmin$^{-1}$ at altitude of 20\,deg.
In any case, in order to minimize any higher order distortions,
it is recommended to divide the image to small sections (say 10 by 10 arcmin).

The typical amplitude of differential aberration of light (due to the Earth motion) is of the order
of $\sim0.2''$\,deg$^{-1}$. This is small enough to be ignored in some cases.
However, since the effect of aberration is fully predictable
it is straightforward to incorporate it into the transformation.
As far as we know popular image registration
(and astrometric) packages
ignore the aberration of light.


\subsection{Astrometric scintillations}
\label{sec:scint}

Astrometric registration of ground-based imaging
is typically limited by astrometric scintillation
induced by the Earth atmosphere.
An order of magnitude estimate for the amplitude of astrometric scintillation is:
\begin{align}
\sigma_{{\rm scint}} \sim \frac{FWHM}{\sqrt{t_{{\rm int}}/t_{{\rm scint}} }},
\label{eq:sigmapos}
\end{align}
where $FWHM$ is the PSF FWHM,
$t_{{\rm int}}$ is the integration time,
and $t_{{\rm scint}}$ is the correlation time scale of
the tip/tilt term of the atmospheric scintillations.
For example, assuming $FWHM=2''$, $t_{{\rm int}}=60$\,s,
and $t_{{\rm scint}}=0.03$\,s,
we get $\sigma_{{\rm scint}}\sim 40$\,mas.
This can be an order of magnitude larger than the astrometric noise
induced by the Poisson noise of bright stars.
In practice this noise depends on the angular scale
(see e.g., \citealp{Shao1992}).

This kind of astrometric noise is hard to remove,
and therefore we expect that bright stars will always
have some leftover residuals in the subtraction process.
However, we presented two methods to deal with this
problem in \S\ref{sec:SourceNoise} and \S\ref{sec:AstNoise}.

\subsection{Color refraction}

\begin{figure}
\centerline{\includegraphics[width=8.5cm]{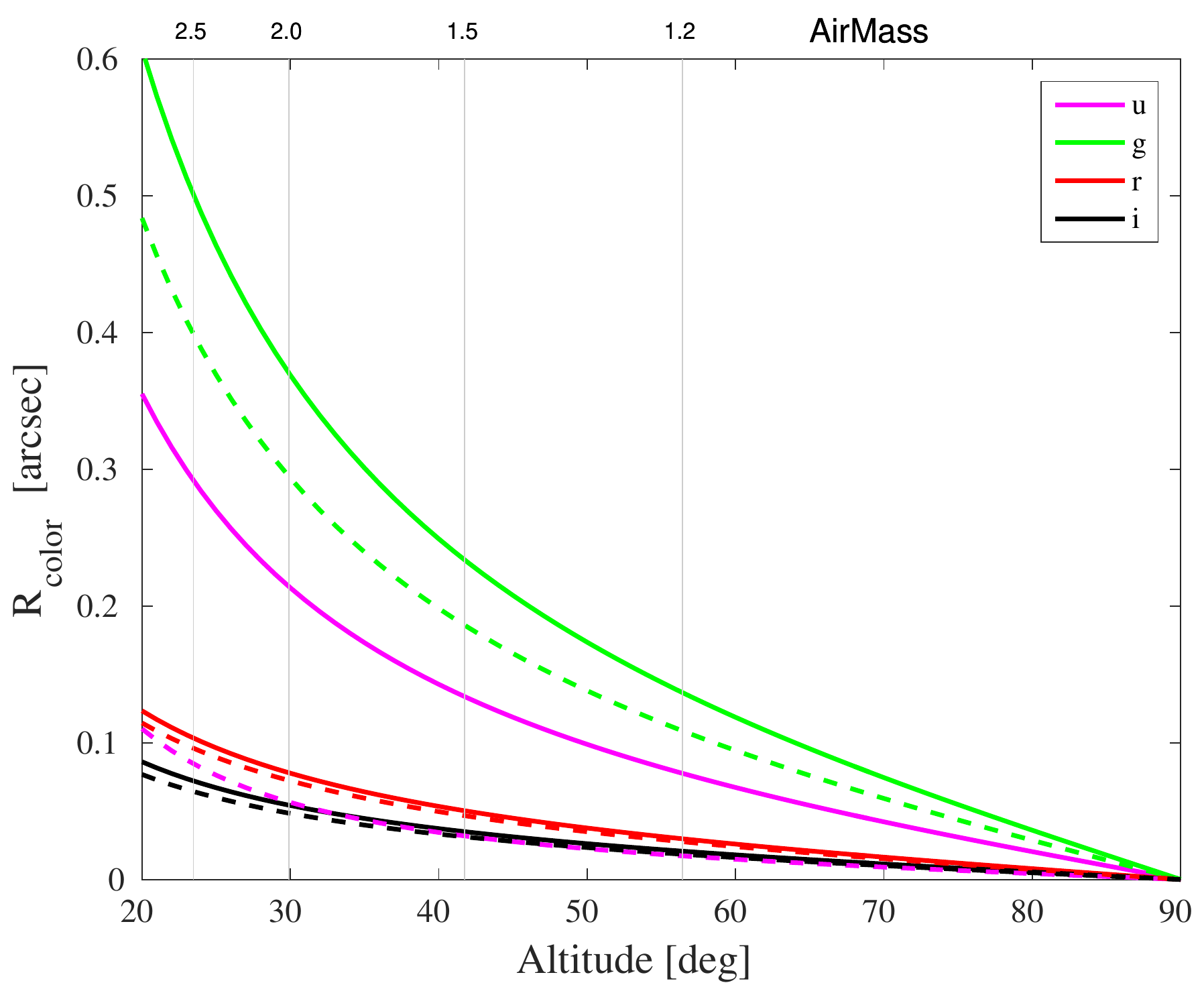}}
\caption{Solid lines represent the difference in color refraction (in the altitude direction) in arcsec,
as a function of altitude, between an O5V star and M5V star.
The various colors correspond to different filters (see legend).
The dashed lines show the same, but for the difference between an A0V star and M5V star.
The calculation includes the atmospheric extinction (at Kitt Peak),
and uses stellar spectra (adopted from Pickles 1998).
Atmospheric conditions are the same as in Figure~\ref{fig:DiffRef_Alt}.
\label{fig:ColorRef_Alt}}
\end{figure}

The atmospheric refraction is color dependent and hence
sources with different spectra will suffer different refraction
at the same airmass.
Figure~\ref{fig:ColorRef_Alt} presents the relative amplitude of
color refraction, in different bands, between an O5V star and an M5V star
and between an A0V star and an M5V star,
as a function of altitude.
%

We suggest three solutions to this issue:
(1) Construct reference images
for several airmass ranges.
Since color refraction is symmetric around the meridian,
one needs to construct such reference images separately for
observations conducted east and west of the meridian;
(2) Calculate the variance induced by this effect and introduce it
as extra terms in the denominator of $S_{{\rm corr}}$ (Eq.~\ref{eq:Scorr});
(3) Fit the astrometric shift for each residual in $D$ using
the scheme presented in \S\ref{sec:AstNoise}.
The last option is likely the best approach.

\subsection{Additional sources of noise}
\label{sec:Add}

There may be additional sources of noise that can influence
image subtraction.
An example for a rare problem we encountered in our simulations
and real images
is that if a binary star has uncorrelated astrometric
noise\footnote{In reality this is rare as both registration
errors and astrometric scintillation noise (but not the
Poisson noise) are correlated on short
angular scales.}
this may affect the calculation of the gradient image
(Eqs.~\ref{eq:VastSN}--\ref{eq:VastSR}; see Figure~\ref{fig:Test1a} for example).
In principle such problems can be accounted for in $S_{{\rm corr}}$,
however, one needs to identify these issues.
Therefore, successful implementation of this method
requires large-scale tests on real data. Such tests are underway,
and this may be further discussed in future publications.

\section{Summary}
\label{sec:summary}

Current popular image subtraction methods have several important limitations,
including: non-optimality, numerical instability in some cases,
some of the methods use matrix inversion which is slow to calculate.
Most importantly, these methods do~not provide
a closed form formula for calculation of the significance
of a transient candidate.
Moreover, in some cases due to numerical instability of some of the methods
it is not possible to calculate, even numerically, the significance of a transient candidate.
This undermines any automatic transient detection and classification,
and may be a considerable obstacle for future surveys.

We present closed-form transient detection and image subtraction statistics that potentially solve all of the above problems, and have the following properties:
\begin{enumerate}
\item The transient detection statistic is mathematically proven to be optimal in the background-dominated noise limit;
\item Both statistics are numerically stable for any pair of input images;
\item For accurately registered, adequately sampled images, these statistics do~not leave any subtraction residuals or deconvolution artifacts;
\item It is possible to correct the transient detection statistic to be resilient to registration errors, color-refraction errors, and any noise for which a model can be constructed;
\item We can assign credible detection significance for newly found transients;
\item The proper subtraction image has white noise in the background-dominated-noise limit. This makes it attractive for more complex measurements and visualization;
\item The proper subtraction statistic is a sufficient statistic for any further statistical test on the difference image. In particular, it allows to distinguish particle hits and other image artifacts from real transients;
\item Both statistics are symmetric to the exchange of the new and reference images;
\item Both statistics are fast to calculate - at least an order of magnitude faster to compute than popular methods;
\item Both statistics are given in closed form and they are straightforward to implement;
\item The proper subtraction statistic allows to search for small astrometric changes between the new and reference images, even in arbitrarily crowded regions; 
\item The same statistics are also optimal for flux measurements in the background-noise dominated limit;
\item We show that the optimal way to prepare a reference image is the proper image coaddition statistic
presented in Zackay \& Ofek (2015b).
\end{enumerate}

We demonstrate this method on simulated data and real observations from the Palomar Transient Factory data release 2.
A summary of the algorithm and equations are presented in \S\ref{sec:algo},
while a discussion regarding the implementation is in \S\ref{sec:Details}.
We briefly describe our MATLAB and Python code that implement
this method and are available online.

We conclude that this image differencing algorithm has the potential
to solve most of the challenges of astronomical image subtraction.
However, testing if this method is indeed capable of completely removing
the need for post-subtraction processing (e.g., human scanners)
requires considerable research effort and tests on large datasets.
Such tests are underway.

We thank Ora Zackay and Assaf Horesh for many discussions.
This paper is based on observations obtained with the
Samuel Oschin Telescope as part of the Palomar Transient Factory
project, a scientific collaboration between the
California Institute of Technology,
Columbia University,
Las Cumbres Observatory,
the Lawrence Berkeley National Laboratory,
the National Energy Research Scientific Computing Center,
the University of Oxford, and the Weizmann Institute of Science.
B.Z. is grateful for receiving the Clore fellowship.
E.O.O. is incumbent of
the Arye Dissentshik career development chair and
is grateful for support by
grants from the 
Willner Family Leadership Institute
Ilan Gluzman (Secaucus NJ),
Israel Science Foundation,
Minerva, Weizmann-UK,
and the I-Core program by the Israeli Committee for Planning
and Budgeting and the Israel Science Foundation (ISF).
A.G. acknowledges support from the I-Core program "The Quantum Universe",
as well as from the Kimmel Award.

\appendix

\section{Full derivation of the image subtraction statistics}
\label{App:FullDeriv}

Let $R$ and $N$ be the background subtracted reference image and background subtracted new image, respectively.
Denote by $T$ the background subtracted true constant sky image.
Denote by $P_r$ and $P_n$ the point spread functions (PSFs)  of the reference image and the new image, respectively.
$P_r$ and $P_n$ are normalized to have unit sum.

Writing the expression for the reference image:
\begin{equation}
R = F_{r} T\otimes P_r + \epsilon_r,
\label{eqA:Rdef}
\end{equation}
where $\epsilon_r$ is the additive noise component of the image $R$.
Given the null hypothesis,  $\mathcal{H}_0$, that states there are no new sources in the new image we can write:
\begin{equation}
N_{|\mathcal{H}_0} = F_{n} T\otimes P_n + \epsilon_n.
\label{eqA:NdefH0}
\end{equation}
Given the alternative hypothesis, $\mathcal{H}_1(q,\alpha)$, that states there is a new point source at position $q$ with flux $\alpha$, we can write:

\begin{equation}
N_{|\mathcal{H}_1(q,\alpha)} = F_{n} T\otimes P_n + F_{n}\alpha\delta(q)\otimes P_n  + \epsilon_n \,,
\label{eqA:NH1}
\end{equation}
where $\delta(q)$ denotes a two dimensional image with one at position $q$, and zero otherwise.
Assuming that the images are background subtracted, and that the dominant source of noise is the background noise,
$\epsilon_r$ and $\epsilon_n$ both satisfy that all pairs of pixels are uncorrelated -- i.e., that for all pairs of pixels
$x_1,x_2$ for which $x_1\neq x_2$:
\begin{align}
{\rm Cov}\left(\epsilon_r[x_1],\epsilon_r[x_2]\right) = 0 \,,{\rm Cov}\left(\epsilon_n[x_1],\epsilon_n[x_2]\right) = 0,
\end{align}
and that all pixels have spatially uniform variance\footnote{In practice this assumption can be relaxed.}:
\begin{equation}
V(\epsilon_r[x]) = \sigma_r^2\,,V(\epsilon_n[x]) = \sigma_n^2.
\end{equation}

Because both hypotheses are simple, we can use the Neyman-Pearson lemma \citep{NeymanPearsonLemma},
that states the most powerful statistic for deciding between two simple hypotheses is the likelihood ratio test:
\begin{align}
\mathcal{L}(q,\alpha) = \frac{\mathcal{P}(N,R|\mathcal{H}_0)}{\mathcal{P}(N,R|\mathcal{H}_1(q,\alpha))},
\end{align}
where $\mathcal{P}$ denotes probability.
A critical point is that we do not have any prior information or assumptions on $T$. Therefore, we cannot calculate the probabilities
$\mathcal{P}(N,R|\mathcal{H}_0)$ and $\mathcal{P}(N,R|\mathcal{H}_1(q,\alpha))$ directly.
However, we can calculate their ratio by developing the expression using the law of conditional probabilities
\begin{align}
\mathcal{L}(q,\alpha) = \frac{\mathcal{P}(N|R,\mathcal{H}_0)\mathcal{P}(R|\mathcal{H}_0)}{\mathcal{P}(N|R,\mathcal{H}_1(q,\alpha))\mathcal{P}(R|\mathcal{H}_1(q,\alpha))}\,.
\end{align}
Using the fact that both $\mathcal{H}_0$ and $\mathcal{H}_1(q,\alpha)$ state the same probabilistic model for $R$ (and therefore will assign the same likelihood for observing $R$) we can further simplify:
\begin{align}
\mathcal{L}(q,\alpha) = \frac{\mathcal{P}(N|R,\mathcal{H}_0)}{\mathcal{P}(N|R,\mathcal{H}_1(q,\alpha))}\,.
\end{align}

To calculate $\mathcal{P}(N|R,\mathcal{H}_0)$ we examine the statistical behavior of the
Fourier transforms of $N$ and $R$ given both hypotheses,
and assume that the images are background-noise dominated.
Using the fact that the Fourier transform of white noise is itself white noise, we know the exact noise properties of the Fourier transform of both $R,N$ given both hypotheses:
\begin{align}
\widehat{N}_{|\mathcal{H}_0} = F_{n}\widehat{T}\widehat{P_n} + \widehat{\epsilon_n},
\label{eqA:N|H_0}
\end{align}
\begin{align}
\widehat{N}_{|\mathcal{H}_1(q,\alpha)} = F_{n}(\widehat{T} + \alpha\widehat{\delta(q)})\widehat{P_n} + \widehat{\epsilon_n},
\label{eqA:N|H_1}
\end{align}
\begin{align}
\widehat{R}_{|\mathcal{H}_0} = \widehat{R}_{|\mathcal{H}_1(q,\alpha)} = F_{r} \widehat{T}\widehat{P_r} + \widehat{\epsilon_r},
\label{eqA:R|H_0}
\end{align}
where the $\widehat{\quad}$ accent denotes Fourier transform and both $\widehat{\epsilon_n}$ and $\widehat{\epsilon_r}$ are complex white
Gaussian noise\footnote{The noise in the Fourier transform of an image with white noise, is white except for the obvious symmetry $\widehat{\epsilon_n}(f_1,f_2) = \overline{\widehat{\epsilon_n}(-f_1,-f_2)}$, where over line denotes complex conjugation. This symmetry is due to the fact that the input images are real.}.

Using the fact that $\widehat{R}$ is measured,
we can invert its probabilistic
model to obtain a model for $T$:
\begin{align}
\widehat{T} = \frac{\widehat{R}}{F_{r} \widehat{P_r}} - \frac{\widehat{\epsilon_r}}{F_{r} \widehat{P_r}}.
\label{eqA:Tr}
\end{align}
Using this expression for $\widehat{T}$, we can write a probabilistic model for $N$ given $R$ and $\mathcal{H}_{0}$:
\begin{align}
\widehat{N}_{|\widehat{R}} = \frac{\widehat{R}}{F_{r} \widehat{P_r}} F_{n}\widehat{P_n} - \frac{\widehat{\epsilon_r}}{F_{r} \widehat{P_r}} F_{n}\widehat{P_n} + \widehat{\epsilon_n} \,.
\end{align}
Given this model for $T$ and assuming the noise is Gaussian,
we can calculate the probability to observe $N$ (this is the $\chi^{2}$ up to a factor of 2):
\begin{align}
\log(\mathcal{P}[\widehat{N}|\widehat{R},\mathcal{H}_0]) = \sum_{f}\frac{\left|\widehat{N} - \frac{ F_{n}\widehat{P_n}\widehat{R}}{F_{r} \widehat{P_r}}\right|^2}{2V(\widehat{\epsilon_n} + \frac{F_{n}\widehat{P_n}\widehat{\epsilon_r}}{F_{r} \widehat{P_r}})}\,,
\end{align}
Using the linearity and scalar multiplication properties of the variance and simplifying we get:
\begin{align}
\log(\mathcal{P}[\widehat{N}|\widehat{R},\mathcal{H}_0]) = \frac{1}{2}\sum_{f}\frac{\left|F_{r} \widehat{P_r}\widehat{N} - F_{n}\widehat{P_n}\widehat{R}\right|^2}{\sigma_n^2 F_{r}^{2}|\widehat{P_r}|^2 + \sigma_r^2 F_{n}^{2} |\widehat{P_n}|^2},
\label{eqA:NhatGivenRhatH0}
\end{align}
Similarly, given $\mathcal{H}_1$ we can write:
\begin{align}
\log(&\mathcal{P}[\widehat{N}|\widehat{R},\mathcal{H}_1(q,\alpha)]) =\\ &=\frac{1}{2}\sum_{f}\frac{\left| F_{r} {\widehat{P_r}}{\widehat{N}} - F_{n} {\widehat{P_n}}{\widehat{R}} - \alpha F_{n}F_{r} {\widehat{P_n}\widehat{P_r}}{\widehat{\delta(q)}}\right|^2}{\sigma_n^2 F_{r}^{2}|\widehat{P_r}|^2 + \sigma_r^2 F_{n}^{2} |\widehat{P_n}|^2}.
\label{eqA:NhatGivenRhatH1}
\end{align}
Now, we can express the log-likelihood ratio test statistic by subtracting Equation \ref{eqA:NhatGivenRhatH1} from Equation \ref{eqA:NhatGivenRhatH0},
opening the absolute value squared using $|a+b|^2 = |a|^2 + |b|^2 + 2\Re[a\overline{b}]$, where $\Re$ is the real number operator,
and removing parts that do not depend on the data:
\begin{align}
&\log(\mathcal{L}(q,\alpha)) =\\ &= \sum_{f}\frac{\Re\left[\left({F_r\widehat{P_r}{\widehat{N}}} - {F_n\widehat{P_n}}{\widehat{R}}\right) \overline{ \alpha{F_rF_n\widehat{P_n}\widehat{P_r}}{\widehat{\delta(q)}}}\right]}{\sigma_n^2F_r^2|\widehat{P_r}|^2 + \sigma_r^2 F_n^2|\widehat{P_n}|^2}.
\end{align}
Noticing that $\alpha$ enters only as a scalar multiplier to the entire expression, we can define a statistic
\begin{align}
S(q)\equiv  \frac{\log(\mathcal{L}(q,\alpha))}{\alpha},
\end{align}
to test optimally for all values of $\alpha$ simultaneously. 

In order to express the same score in term of intuitive quantities we define the {\it proper subtraction} image:
\begin{align}
\widehat{D} = \frac{\left(F_r{\widehat{P_r}}{\widehat{N}} - F_n{\widehat{P_n}}{\widehat{R}}\right)}{\sqrt{\sigma_n^2F_r^2|\widehat{P_r}|^2 + \sigma_r^2F_n^2|\widehat{P_n}|^2}}.
\label{eqA:D_hat}
\end{align}
The PSF for transient detection:
\begin{align}
\widehat{P_D} = \frac{F_rF_n\widehat{P_r}\widehat{P_n}}{F_D\sqrt{\sigma_n^2F_r^2|\widehat{P_r}|^2 + \sigma_r^2F_n^2|\widehat{P_n}|^2}}\,,
\end{align}
and the normalization:
\begin{align}
F_D = \frac{F_nF_r}{\sqrt{F_n\sigma_r^2 + F_r\sigma_n^2}}.
\end{align}
We note that $F_{D}$ can be derived by substituting 1 into $\widehat{P_{n}}$ and $\widehat{P_{r}}$ in the expression for $\widehat{P_{D}}$.

In the background-noise dominated limit, $D$ has white noise (see \S\ref{sec:whitenoise}).
The score $S(q)$ can now be expressed by: 
\begin{align}
S(q) = \Re\left[F_D\sum_{f}{\widehat{D}\overline{\widehat{P_D}\widehat{\delta(q)}}}\right]\,.
\end{align}
Expressing this in real space using the convolution theorem we get:
\begin{align} S(q) = F_D\Re\left[D \otimes\overleftarrow{P_D}\otimes\overleftarrow{\delta(q)}\right](0)\,.
\end{align}
Noticing that both $D$ and $P_D$ contain only real numbers, the real operator can be removed. Convolution with a delta function is just the shift operator, therefore the expression for $S(q)$ can be simplified even further to be:
\begin{align} S(q) = [F_D D \otimes\overleftarrow{P_D}](q).
\end{align}
The expression for its Fourier transform is then expressed by:
\begin{align}
\widehat{S} = \widehat{D}\overline{\widehat{P_{D}}} = \frac{F_nF_r^2\overline{\widehat{P_n}}|\widehat{P_r}|^2\widehat{N} - F_rF_n^2\overline{\widehat{P_r}}|\widehat{P_n}|^2\widehat{R} }{\sigma_r^2F_n^2|\widehat{P_n}|^2 + \sigma_n^2F_r^2|\widehat{P_r}|^2}.
\label{eqA:S}
\end{align}
This is the final form of the optimal transient detection statistic. 
An alternative form for this expression can be written as:
\begin{align}
\widehat{S} = \frac{F_nF_r^2\frac{\overline{\widehat{P_n}}}{\sigma_n^2}\frac{|\widehat{P_r}|^2}{\sigma_r^2}\widehat{N} - F_rF_n^2\frac{\overline{\widehat{P_r}}}{\sigma_r^2}\frac{|\widehat{P_n}|^2}{\sigma_n^2}\widehat{R} }{F_n^2\frac{|\widehat{P_n}|^2}{\sigma_n^2} + F_r^2\frac{|\widehat{P_r}|^2}{\sigma_r^2}}\,.
\label{eqA:Salt}
\end{align}

\section{Construction of the reference image}
\label{secA:RefIm}

Extending the statistical framework to the situations in which we are given a set of references,
we seek to find the optimal transient detection statistic given all of the references.
Each reference image out of a total of $J$ images is given by:
\begin{align}
R_j = F_jP_j\otimes T + \epsilon_j.
\end{align}
A certain new image $N$ is measured, and we want to determine which of the following is true,
$\mathcal{H}_0$:
\begin{align}
N = F_nP_n\otimes T + \epsilon_n,
\end{align}
or $\mathcal{H}_1(q)$:
\begin{align}
N = F_nP_n\otimes (T + \delta(q)) + \epsilon_n.
\end{align}

As in the previous section, we are trying to test between two simple hypotheses. Therefore, the optimal test statistic is the log-likelihood ratio test \citep{NeymanPearsonLemma}
\begin{align}
\mathcal{L}(q,\alpha) = \frac{\mathcal{P}(N,R_1,\dots,R_J|\mathcal{H}_0)}{\mathcal{P}(N,R_1,\dots,R_J|\mathcal{H}_1(q,\alpha))}.
\end{align}
As before, we can use the law of conditional probabilities, and the fact that $\mathcal{H}_0$ and $\mathcal{H}_1$ predict the same likelihood to all references:
\begin{align}
\mathcal{L}(q,\alpha) = \frac{\mathcal{P}(N|R_1,\dots,R_J,\mathcal{H}_0)}{\mathcal{P}(N|R_1,\dots,R_J,\mathcal{H}_1(q,\alpha))}.
\end{align}
In order to calculate the conditional probabilities, we need a probabilistic model for $N$ that does not contain $T$. This could be achieved by using all references to get the best statistical model for $T$.

As in the previous section, this can be more easily formulated by stating the hypotheses
for the images in the Fourier plane:
\begin{align}
\widehat{N}_{|\mathcal{H}_0} = \widehat{T}\widehat{P_n} + \widehat{\epsilon_n},
\label{eqA:N_multi|H_0}
\end{align}
\begin{align}
\widehat{N}_{|\mathcal{H}_1(q,\alpha)} = (\widehat{T} + 
\alpha\widehat{\delta(q)})\widehat{P_n} + \widehat{\epsilon_n},
\label{eqA:N_multi|H_1}
\end{align}
\begin{align}
\widehat{R_j}_{|\mathcal{H}_0} = \widehat{R_j}_{|\mathcal{H}_1(q,\alpha)} = \widehat{T}\widehat{P_j} + \widehat{\epsilon_j}. 
\label{eqA:R_multi|H_0}
\end{align}

Following Appendix~\ref{App:FullDeriv}, 
we can continue to develop this in the long way into the correct
difference image and the correct transient detection statistic.
However, we can take a shortcut.
The key observation we make, is that we can cast all the information in the reference images into a statistical model for $\widehat{T}$.
Using the result from the appendix of Zackay and Ofek 2015a
(paper~I in the series on coaddition), the choice that maximizes the $S/N$
is the weighted addition of all the sources of information on $\widehat{T}(f)$:
\begin{align}
\widehat{T} = \frac{\sum_j\frac{F_j\overline{\widehat{P}_j}}{\sigma_j^2}\widehat{R_j}}{\sum_j\frac{F_j^2|\widehat{P}_j|^2}{\sigma_j^2}} + \widehat{\epsilon_T}.
\end{align}
Where we have denoted the noise contribution from all the reference images by $\widehat{\epsilon_T}$.
Calculating its variance we get that:
\begin{align}
V[\widehat{\epsilon_T}] = \frac{1}{\sum_j\frac{F_j^2|\widehat{P}_j|^2}{\sigma_j^2}} \equiv \frac{1}{F_r^2|\widehat{P_r}|^2},
\end{align}
where we have defined: 
\begin{align}
F_r = \sqrt{\sum_j{\frac{F_j^2}{\sigma_j^2}}}\;,\quad \widehat{P_r} = \frac{1}{F_r}\sqrt{\sum_j{\frac{F_j^2}{\sigma_j^2}|\widehat{P_j}|^2}}.
\end{align}
Given these choices and the template of Equation~\ref{eqA:Tr},
we find the formula for the coaddition of the reference images:
\begin{align}
\widehat{R} = \frac{\sum_j F_{j}\frac{\overline{\widehat{P}_j}}{\sigma_j^2}\widehat{R_j}}{\sqrt{\sum_j F_{j}^{2}\frac{|\widehat{P}_j|^2}{\sigma_j^2}}}.
\label{eqA:Rprop}
\end{align}
Here $\widehat{\sigma_R} = 1$.
Since $R$, $P_{r}$ and $T$ satisfies Equation~\ref{eqA:Rdef},
we have a single reference image that complies with the requirements of the statistical model.
Interestingly, Equation~\ref{eqA:Rprop} is identical to the
proper coaddition image presented in Zackay and Ofek (2015b; paper~II in the series of coaddition).

Substituting Equation~\ref{eqA:Rprop} into $\widehat{D}$ we get:
%
\begin{align}
\widehat{D} = \frac{{\sqrt{\sum_j{\frac{F_j^2|P_j|^2}{\sigma_j^2}}}}{\widehat{N}} - {F_n\widehat{P_n}}\left(\frac{\sum_j\frac{\overline{F_j\widehat{P}_j}}{\sigma_j^2}\widehat{R_j}}{\sqrt{\sum_j\frac{F_j^2|\widehat{P}_j|^2}{\sigma_j^2}}}\right)}{\sqrt{\sigma_n^2\left(\sum_j{\frac{F_j^2|P_j|^2}{\sigma_j^2}}\right) + \sigma_r^2F_n^2|\widehat{P_n}|^2}}.
\label{eq:D_hat_multi}
\end{align}
Writing the source detection statistic in explicit form we get:
\begin{align}
\widehat{S} = \frac{\frac{F_n\overline{\widehat{P_n}}}{\sigma_n^2}\left(\sum_{j}{\frac{F_j^2|\widehat{P}_j|^2}{\sigma_j^2}}\right)\widehat{N} - \frac{F_n^2|\widehat{P_n}|^2}{\sigma_n^2}\left(\sum_{j}{\frac{F_j\overline{\widehat{P}_j}\widehat{R}_j}{\sigma_j^2}}\right) }{\frac{F_n^2|\widehat{P_n}|^2}{\sigma_n^2} + \left(\sum_{j}{\frac{F_j^2|\widehat{P}_j|^2}{\sigma_j^2}}\right)}.
\label{eq:Sstat}
\end{align}

Thus, we arrive at an optimal solution with a closed formula for optimal transient detection given a set of references.
We note that there are other choices that can be used instead of $R$. However, we prefer the proper coaddition image
due to its uncorrelated noise (see Zackay \& Ofek 2015b).
Finally, also $N$ can be composed of multiple images.
In this case, the optimal solution for the subtraction is to perform the optimal transient detection with both $N$ and $R$
being the proper coaddition of all the images in their corresponding sets.

\section {Correction for source noise of bright objects}
\label{Ap:CorrectSourceNoise}

The assumption that the noise distribution is independent of position, and of the true image itself, is of course not true. Specifically, near bright stars the dominant source of noise is the Poisson fluctuations of the source itself, which is obviously position dependent.
Therefore, in the vicinity of bright sources the variance is underestimated,
and random fluctuations in the noise can cause false transient detections in these positions.
Since only a negligible part of the sky behaves in such a way, we do not wish to change the statistic $S$ in places away from bright sources.

Therefore, the approach we currently recommend is the following:
Calculate separately the two parts of Equation~\ref{eq:Sstat}:
\begin{align}
\widehat{S_N}  = \frac{\frac{F_n\overline{\widehat{P_n}}}{\sigma_n^2}\left(\sum_{j}{\frac{F_j^2|\widehat{P}_j|^2}{\sigma_j^2}}\right) }{\frac{F_n^2|\widehat{P_n}|^2}{\sigma_n^2} + \left(\sum_{j}{\frac{F_j^2|\widehat{P}_j|^2}{\sigma_j^2}}\right)}\widehat{N} \equiv \widehat{k_n} \widehat{N},
\label{eqA:SN}
\end{align}
and
\begin{align}
\widehat{S_{R_j}} = \frac{ \frac{F_n^2|\widehat{P_n}|^2}{\sigma_n^2}\frac{F_j\overline{\widehat{P}_j}}{\sigma_j^2} }{\frac{F_n^2|\widehat{P_n}|^2}{\sigma_n^2} + \left(\sum_{j}{\frac{F_j^2|\widehat{P}_j|^2}{\sigma_j^2}}\right)}\widehat{R}_j \equiv \widehat{k_j}\widehat{R_j}.
\label{eqA:SR}
\end{align}
Next, apply inverse Fourier transform to get to the image domain:
\begin{align}
S = S_N - \sum_j{S_{R_j}}.
\end{align}
Then calculate the corrected score for the existence of transient sources:
\begin{align}
S_{{\rm corr}} = \frac{S_N - \sum_j{S_{R_j}}}{\sqrt{V(S_N) + \sum_j{V(S_{R_j})}}},
\label{eqA:Scorr}
\end{align}
where $V(S_N)$ and $V(S_{R_j})$ are the variance maps of $S_N$ and $S_{R_j}$.
Essentially, these can be computed analytically by following all the operations done on $R_j$ and $N$,
and applying the corresponding corrections to $V(S_{R_j})$ and $V(S_N)$ respectively. 

Using the fact that for a zero expectancy noise source $\epsilon$, 
\begin{align}
{\rm V}(\epsilon\otimes P) = V(\epsilon)\otimes (P^2)\,.
\end{align}
we can derive a closed formula solution for $V(S_N)$ and $V(S_{R_j})$:
\begin{align}
V(S_N) = V(\epsilon_n)\otimes(k_n^2)\,,
\end{align} 
\begin{align}
V(S_{R_j}) = V(\epsilon_j)\otimes(k_j^2),
\label{eqA:VSRj}
\end{align} 
where $k_n$ and $k_j$ are defined in Equations~\ref{eqA:SN}
and \ref{eqA:SR}, respectively.
We note that the squaring of the convolution kernel happens in the image domain.

In the presence of bright stars the noise
is correlated, this means that we need to store,
or sum up the individual $V(S_{R_j})$.
Using the proper coaddition image and its effective $k_{r}$
will not recover all the information.
However, using $R$ and $k_{r}$ may serve as an approximation
to this process.

%
%

The proposed correction (Eq.~\ref{eqA:Scorr})
does not change the score image away from bright sources
(other than move the detection statistic to units of standard deviations).
The reason for this is that the variance map is spatially uniform
in places away from strong sources.
We note that this correction is sub-optimal near bright sources,
but at least it is a score with known statistical properties,
that we can use to prevent false positives and to retain some sensitivity.

This method of correcting $S$ by the variance can be extended to any additional sources of noise
for which we can construct a model.
For example, in \S\ref{sec:SourceNoise} we present also the variance due to astrometric errors.

\section{Optimal PSF photometry of transient point sources}
\label{Ap:optimalPhotometry}


In general, in the statistical community, there is no consensus
on how to derive the best measurement.
Therefore, in this section, we will search for a measurement
statistic that is unbiased and has maximal $S/N$,
and is a linear function of the input images.
Not surprisingly, the resulting statistics is simply
$S$ (Equation \ref{eq:S}) normalized by some factor.
This analysis also presents another formalism in which our
transient detection statistic is optimal -- it is the maximum $S/N$
linear statistic composed out of $R$ and $N$ that cancels the constant in time image $T$.

We start by stating again the statistical model we use:
\begin{align}
R &= P_r\otimes T + \epsilon_r,\\
N &= P_n\otimes (T + \alpha\delta(q)) + \epsilon_n,
\end{align}
where $\alpha$ is the flux of the new source at
position $q$, and $\delta(q)$ is an image with $0$
everywhere except position $q$ where it's value is $1$.
We continue to work under the assumption that the
background noise is the most significant source of noise,
which allows us to write:
\begin{align}
V[\epsilon_r] = \sigma_r^2\,, \quad V[\epsilon_n] = \sigma_n^2.
\end{align}

We write the statistic that we are looking for in its most general linear form:
\begin{align}
C = k_n\otimes N + k_r\otimes R,
\end{align}
where $k_{n}$ and $k_{r}$ are some kernels,
and we require that:
\begin{align}
F_n k_n\otimes P_n = -F_rk_r\otimes P_r.
\end{align}
Writing $C$ in Fourier space we get:
\begin{align}
\widehat{C} = \widehat{k_n}\widehat{N} + \widehat{k_r}\widehat{R} = \alpha\widehat{\delta(q)}F_n\widehat{P_n}\widehat{k_n} + \widehat{\epsilon_c},
\end{align}
where $\epsilon_c$ absorbs all noise sources in both images.

Here, we will use a well known result
(also given in Appendix~B of Zackay \& Ofek 2015a) that the maximal $S/N$
measurement of a parameter $\theta$ given a set of statistics $X_j$ such that:
\begin{align}
X_j = \mu_j\theta +\epsilon_j,
\end{align}
where $\mu_j$ are scaling factors and $\epsilon_j$ has variance $V[\epsilon_j] = \sigma_j$, is:
\begin{align}
\widetilde{\theta} = \frac{\sum_j{\frac{\overline{\mu_j}}{\sigma_j^2}X_j}}{\sum_j{\frac{|\mu_j|^2}{\sigma_j^2}}}.
\end{align}

In our case $\mu=\widehat{\delta(q)}F_n\widehat{P_n}\widehat{k_n}$.
Applying this to $\widehat{C}$, we get the maximum $S/N$ statistic for $\alpha$:
\begin{align}
\widetilde{\alpha} &=  \frac{\sum_{f}{\frac{\overline{\widehat{\delta(q)}F_n\widehat{P_n}\widehat{k_n}}\widehat{C}}{\sigma_r^2|\widehat{k_r}|^2 + \sigma_n^2|\widehat{k_n}|^2}}} { \sum_{f}{\frac{|{\widehat{\delta(q)}F_n\widehat{P_n}\widehat{k_n}}|^2}{\sigma_r^2|\widehat{k_r}|^2 + \sigma_n^2|\widehat{k_n}|^2}}},
\end{align}
substituting $\widehat{C} = \widehat{k_n}\widehat{N} + \widehat{k_r}\widehat{R}$,
and
\begin{align}
\widehat{k_r}=-\widehat{k_n}\frac{F_n\widehat{P_n}}{F_r\widehat{P_r}},
\end{align}
and simplifying (notice the cancellation of $k_n$ in the ratio, and the use of $|\widehat{\delta(q)}|=1$) we get:
\begin{align}
\widetilde{\alpha} &=\frac{\sum_{f}{\frac{\overline{\widehat{\delta(q)}F_n\widehat{P_n}\widehat{k_n}}(\widehat{k_n}\widehat{N} + \widehat{k_r}\widehat{R})}{\sigma_r^2|\widehat{k_r}|^2 + \sigma_n^2|\widehat{k_n}|^2}}} { \sum_{f}{\frac{|{\widehat{\delta(q)}F_n\widehat{P_n}\widehat{K_n}}|^2}{\sigma_r^2|\widehat{k_r}|^2 + \sigma_n^2|\widehat{k_n}|^2}}} =
\frac{\sum_{f}{\frac{\overline{\widehat{\delta(q)}F_n\widehat{P_n}}(\widehat{N} - \frac{F_n\widehat{P_n}}{F_r\widehat{P_r}}\widehat{R})}{\sigma_r^2|\frac{F_n\widehat{P_n}}{F_r\widehat{P_r}}|^2 + \sigma_n^2}}} { \sum_{f}{\frac{F_n^2|\widehat{P_n}|^2}{\sigma_r^2|\frac{F_n^2\widehat{P_n}}{F_r^2\widehat{P_r}}|^2 + \sigma_n^2}}},
\end{align}
\begin{align}
\widetilde{\alpha}=\frac{\sum_{f}{\frac{\overline{\widehat{\delta(q)}}(F_r^2F_n|\widehat{P_r}|^2\overline{\widehat{P_n}}\widehat{N} - F_n^2F_r|\widehat{P_n}|^2\overline{\widehat{P_r}}\widehat{R})}{\sigma_r^2F_n^2|\widehat{P_n}|^2 + \sigma_n^2F_r^2|\widehat{P_r}|^2}}} { \sum_{f}{\frac{F_n^2F_r^2|\widehat{P_n}|^2|\widehat{P_r}|^2}{\sigma_r^2F_n^2|\widehat{P_n}|^2 + \sigma_n^2F_r^2|\widehat{P_r}|^2}}}.
\label{eq:tildeF_pre}
\end{align}

Last, we see that we can calculate all the fluxes for all the transient sources simultaneously by noticing that the numerator in the expression for
$\widetilde{\alpha}$ is the $q$'th position in the previously defined transient detection image $S$ (Equation~\ref{eqA:S}).
That is:
\begin{align}
\widetilde{\alpha}=\frac{S} { \sum_{f}{\frac{F_{n}^{2}F_{r}^{2}|\widehat{P_n}|^2|\widehat{P_r}|^2}{\sigma_r^2 F_{n}^{2}|\widehat{P_n}|^2 + \sigma_n^2 F_{r}^{2}|\widehat{P_r}|^2}}}.
\label{eq:tildeF} 
\end{align}
This means that the same statistic can be computed both for detection and measurement. Therefore, in order to get a flux measurement from $S$,
all we need is to normalize it by $F_S$ -- the denominator of Equation~\ref{eq:tildeF}:

\begin{align}
F_S = \sum_{f}{\frac{F_n^2F_r^2|\widehat{P_n}|^2|\widehat{P_r}|^2}{\sigma_r^2F_n^2|\widehat{P_n}|^2 + \sigma_n^2F_r^2|\widehat{P_r}|^2}}\,.
\end{align}

Via the same process as for the detection, the standard deviation of the flux measurement $S$ at position $q$ can be estimated via inspection of $S_N$ and $S_R$.
We find that the standard deviation of $F$ can be calculated by:
\begin{align}\label{eq:varFluxMeasurement}
\sigma_{\widetilde{\alpha}} = \frac{\sqrt{V(S_N) +{V(S_{R})}}}{F_S}.
\end{align}
If the reference image is constructed from many reference images than
\begin{align}
V(S_R) = \sum_j {V(S_{R_j})}.
\end{align}

Note that Equation \ref{eq:varFluxMeasurement} is valid for both
faint (i.e., in background-dominated-noise areas)
and bright transients (source-dominated-noise areas).
We further note that Equation~\ref{eq:tildeF} is equivalent to PSF photometry
as each pixel is weighted by the appropriate value of the PSF.

\section{$D,P_D,F_D$ are sufficient for any statistical measurement or decision on the difference between the images}
\label{Ap:Sufficiency}

In order to show that $D,P_D,F_D$ are sufficient statistics,
we will use the Fisher-Neyman factorization theorem.
This theorem states that:
If the probability density function is $\mathcal{P}_{\theta}(X)$,
then $T$ is sufficient for
the parameter $\theta$ if and only if nonnegative functions $g$ and $h$
can be found such that
\begin{align}
\mathcal{P}_{\theta}(X) = h(X) g_{\theta}(T(X)).
\label{eq:FNfact}
\end{align}

In our case, we would like to show that 
for any generative model $A_n(\theta)$ for the difference between the images, with parameter $\theta$,
the probability of observing the data ($R$ and $N$) factorizes into:
\begin{align}
\mathcal{P}(R,N|A_n(\theta)) = \mathcal{P}(D|A_n(\theta))g(R,N).
\end{align} 
This will prove that $D$ is a sufficient statistics.

We note that the meaning of sufficient statistics is profound --
it means that any measurement or decision performed on $D$ will return
the same numerical value as if it was performed using all the data.
Examples for such measurements or decisions
are: arbitrary shape measurements, or identifying particle hits.

In this part, we show that $D$, along with $P_D$,$P_{D_N}$,$P_{D_R}$,
are together sufficient for any likelihood calculation
(up to some multiplicative, model independent factor,
as allowed from the Fisher-Neyman criterion)
for any instance of a generative model for $A_n(\theta)$,
regardless of the constant-in-time image $T$.
We state the family of statistical models $A_n(\theta,q)$
for which we want $D$ to be sufficient to:
\begin{align}
R = F_rT\otimes P_r + \epsilon_r,
\end{align}
\begin{align}
N = F_nT\otimes P_n + A_n(\theta)\otimes \delta(q) + \epsilon_n,
\end{align}
where $A_n(\theta)$ is the change made in the new image, located in position $q$, and $T$ is the constant-in-time (unknown) image.
Note that we did~not convolved $A_{n}(\theta)$ with the PSF of the images,
as this will allow us to deal with signal that was not convolved
by the PSF (e.g., bad pixels, small astrometric shifts).
However, such a PSF can be included in $A_{n}(\theta)$.

Using the law of conditional probability, the probability we would like to calculate is:
\begin{align}
\mathcal{P}(R,N|A_n(\theta))&=\mathcal{P}(N|R,A_n(\theta))\mathcal{P}(R|A_n(\theta)) \\&= \mathcal{P}(N|R,A_n(\theta))\mathcal{P}(R).
\end{align}
Since the probability of $R$ is independent of the model
parameter $\theta$ (as it only influences the model for $N$), it suffices for us to calculate $\log(\mathcal{P}(N|R,A_n(\theta)))$.
As we did in previous sections, we can project our knowledge of $R$ to a statistical model for $T$:
\begin{align}
\widehat{T} = \frac{\widehat{R}}{F_{r} \widehat{P_r}} - \frac{\widehat{\epsilon_r}}{F_{r} \widehat{P_r}} \equiv \frac{\widehat{R}}{F_{r} \widehat{P_r}} + \widehat{\epsilon_T}.
\end{align}
We can then use it to calculate the probability of observing $N$ given $A_n(\theta)$:
\begin{align}\label{eq:ProbN|An}
-log(\mathcal{P}(N|R,A_n(\theta))) = \sum_f{\frac{||\widehat{N}-F_n\widehat{P_n}\widehat{T}-\widehat{A_n(\theta)}\widehat{\delta(q)} ||^2}{2V[\widehat{\epsilon_n} + F_n\widehat{P_n}\widehat{\epsilon_T}]}}\,.
\end{align}
Opening the absolute value, we get the summation of three terms. The first term,
$\sum_f{\frac{||\widehat{N}-F_n\widehat{P_n}\widehat{T}||^2}{2V[\widehat{\epsilon_n} + F_n\widehat{P_{n}}\widehat{\epsilon_T}]}}$, does not depend on $A_n(\theta)$ and therefore can be removed
(can be absorbed in the Fisher-Neyman $h$).
The second term is:
\begin{align}
&\sum_f{2\Re\left[\frac{(\widehat{N}-F_n\widehat{P_n}\widehat{T})\overline{\widehat{A_n(\theta)}\widehat{\delta(q)}}}{2V[\widehat{\epsilon_n} + F_n\widehat{P_n}\widehat{\epsilon_T}]}\right]} = ... = \\  &= \frac{(F_r^2|\widehat{P_r}|^2\widehat{N} - F_nF_r\widehat{P_n}\overline{\widehat{P_r}}\widehat{R}) \overline{\widehat{A_n(\theta)}}}{\sigma_n^2F_r^2|\widehat{P_r}|^2 + \sigma_r^2F_n^2|\widehat{P_n}|^2}\,.
\end{align}
In the last expression we can identify a matched filter operation
between the proper subtraction image $D$
\begin{align}
\widehat{D} = \frac{\left(F_r{\widehat{P_r}}{\widehat{N}} - F_n{\widehat{P_n}}{\widehat{R}}\right)}{\sqrt{\sigma_n^2F_r^2|\widehat{P_r}|^2 + \sigma_r^2F_n^2|\widehat{P_n}|^2}},
\end{align}
and the PSF for delta function in $N$ (the PSF of $A_{n}$):
\begin{align}
\widehat{P_{D_N}} = \frac{F_r\widehat{P_r}}{F_{D_N}\sqrt{\sigma_n^2F_r^2|\widehat{P_r}|^2 + \sigma_r^2F_n^2|\widehat{P_n}|^2}},
\label{eqA:Pdn}
\end{align}
with the zero-point:
\begin{align}
F_{D_N} = \frac{F_{r}}{\sqrt{\sigma_{n}^{2}F_{r}^{2} + \sigma_{r}^{2}F_{n}^{2}}}.
\label{eq:FdN}
\end{align}

Finally, we need to show that the third term in Equation \ref{eq:ProbN|An} can be calculated only using $D$ and its set of PSFs and zero-points.
\begin{align}
\sum_{f}\frac{ |\widehat{A_n(\theta)}|^2}{V[\widehat{\epsilon_n} + F_{n}\widehat{P_n}\widehat{\epsilon_T}]} &= \frac{F_r^2|\widehat{P_r}|^2|\widehat{A_n(\theta)}|^2}{\sigma_n^2F_r^2|\widehat{P_r}|^2 + \sigma_r^2F_n^2|\widehat{P_n}|^2} \\ &= F_{D_N}^2|\widehat{A_n(\theta)}|^2|\widehat{P_{D_N}}|^2\,.
\end{align}

Symmetrically, every statistical change in $R$ can be calculated in the same fashion using $D$ and ${P_{D_R}}$.
\begin{align}
\widehat{P_{D_R}} = \frac{F_n\widehat{P_n}}{F_{D_R}\sqrt{\sigma_n^2F_r^2|\widehat{P_r}|^2 + \sigma_r^2F_n^2|\widehat{P_n}|^2}}\,,
\end{align}
with the zero-point 
\begin{align}
F_{D_R} = \frac{F_{n}}{\sqrt{\sigma_{n}^{2}F_{r}^{2} + \sigma_{r}^{2}F_{n}^{2}}}
\label{eq:FdR}
\end{align}
As expected, a change in either $N$ or $R$, that experiences the same PSF (and transparency) as the true image (e.g., a supernovae, variable star or small solar system body) will have the effective PSF $P_D$, and zero-point $F_D$.

This analysis means that the subtraction product $D$, is the optimal statistics for any, even yet unspecified,
measurement or hypothesis testing we wish to perform on the data.


\begin{thebibliography}{}
\expandafter\ifx\csname natexlab\endcsname\relax\def\natexlab#1{#1}\fi


\bibitem[Alard(2000)]{Alard2000} Alard, C.\ 2000, \aaps, 144, 363 

\bibitem[Alard \& Lupton(1998)]{AL98} Alard, C., \& Lupton, R.~H.\ 1998, \apj, 503, 325 

\bibitem[Arcavi et al.(2011)]{Arcavi2011} Arcavi, I., Gal-Yam, A., 
Yaron, O., et al.\ 2011, \apjl, 742, L18 

\bibitem[Baltay et al.(2013)]{LSQ} Baltay, C., Rabinowitz, D., Hadjiyska, E., et al.\ 2013, \pasp, 125, 683

\bibitem[Bellm et al.(2015)]{Bellm2015} Bellm, E.~C., Kulkarni, 
S.~R., 
\& ZTF Collaboration 2015, American Astronomical Society Meeting Abstracts, 225, \#328.04 

\bibitem[Becker et al.(2012)]{Becker2012} Becker, A.~C., Homrighausen, D., Connolly, A.~J., et al.\ 2012, \mnras, 425, 1341 

\bibitem[Bramich(2008)]{Bramich} Bramich, D.~M.\ 2008, \mnras, 
386, L77 

\bibitem[Bramich et al.(2015)]{Bramich2015} Bramich, D.~M., Horne, K., Alsubai, K.~A., et al.\ 2015, arXiv:1512.04655 

\bibitem[Bloom et al.(2012)]{ML1} Bloom, J.~S., Richards, 
J.~W., Nugent, P.~E., et al.\ 2012, \pasp, 124, 1175 

\bibitem[Cenko et al.(2013)]{Cenko2013} Cenko, S.~B., Kulkarni, 
S.~R., Horesh, A., et al.\ 2013, \apj, 769, 130 

\bibitem[Cenko et al.(2015)]{Cenko2015} Cenko, S.~B., Urban, 
A.~L., Perley, D.~A., et al.\ 2015, \apjl, 803, L24 

\bibitem[Gal-Yam et al.(2008)]{Gal-Yam08} Gal-Yam, A., Maoz, D., 
Guhathakurta, P., \& Filippenko, A.~V.\ 2008, \apj, 680, 550 

\bibitem[Gal-Yam et al.(2011)]{Gal-Yam2011} Gal-Yam, A., Kasliwal, M.~M., Arcavi, I., et al.\ 2011, \apj, 736, 159 

\bibitem[Gal-Yam et al.(2014)]{Gal-Yam2014} Gal-Yam, A., Arcavi, I., Ofek, E.~O., et al.\ 2014, \nat, 509, 471

\bibitem[Goldstein et al.(2015)]{ML2} Goldstein, D.~A., 
D'Andrea, C.~B., Fischer, J.~A., et al.\ 2015, arXiv:1504.02936 

\bibitem[Ivezic et al.(2008)]{LSST} Ivezic, Z., Tyson, 
J.~A., Abel, B., et al.\ 2008, arXiv:0805.2366 

\bibitem[Kaiser et al.(2002)]{PANSTARRS} Kaiser, N., Aussel, H., 
Burke, B.~E., et al.\ 2002, \procspie, 4836, 154 

\bibitem[Laher et al.(2014)]{Laher2014} Laher, R.~R., Surace, J., 
Grillmair, C.~J., et al.\ 2014, \pasp, 126, 674 

\bibitem[Law et al.(2009)]{PTF} Law, N.~M., Kulkarni, 
S.~R., Dekany, R.~G., et al.\ 2009, \pasp, 121, 1395

\bibitem[Neyman \& Pearson(1933)]{NeymanPearsonLemma} Neyman, J. \& Pearson, E. S.\ 1933 Philosophical Transactions of the Royal Society of London A: Mathematical, Physical and Engineering Sciences 

\bibitem[Ofek et al.(2011)]{Ofek2011} Ofek, E.~O., Frail, D.~A., 
Breslauer, B., et al.\ 2011, \apj, 740, 65 

\bibitem[Ofek et al.(2012)]{Ofek2012} Ofek, E.~O., Laher, R., 
Law, N., et al.\ 2012, \pasp, 124, 62 

\bibitem[Ofek(2014)]{Ofek2014} Ofek, E.~O.\ 2014, Astrophysics 
Source Code Library, ascl:1407.005 

\bibitem[Ofek et al.(2014)]{Ofeketal2014} Ofek, E.~O., Sullivan, M., 
Shaviv, N.~J., et al.\ 2014, \apj, 789, 104 

\bibitem[Phillips \& Davis(1995)]{Phillips95} Phillips, A.~C., \& Davis, L.~E.\ 1995, Astronomical Data Analysis Software and Systems IV, 77, 297 

\bibitem[Schechter et al.(1993)]{Schechter1993} Schechter, P.~L., 
Mateo, M., \& Saha, A.\ 1993, \pasp, 105, 1342 

\bibitem[Shao 
\& Colavita(1992)]{Shao1992} Shao, M., \& Colavita, M.~M.\ 1992, \aap, 262, 353 

\bibitem[Smith et al.(2011)]{Smith2011} Smith, A.~M., Lynn, S., 
Sullivan, M., et al.\ 2011, \mnras, 412, 1309 

\bibitem[Stetson(1987)]{Stetson1987} Stetson, P.~B.\ 1987, \pasp, 
99, 191 

\bibitem[van Dokkum(2001)]{vanDokkum2011} van Dokkum, P.~G.\ 2001, 
\pasp, 113, 1420 

\bibitem[Wright et al.(2015)]{ML3} Wright, D.~E., Smartt, 
S.~J., Smith, K.~W., et al.\ 2015, \mnras, 449, 451 

\bibitem[Yuan \& Akerlof(2008)]{Yuan08} Yuan, F., \& Akerlof, C.~W.\ 2008, \apj, 677, 808




\end{thebibliography}
\end{document}